\documentclass[11pt,A4paper]{article}
\usepackage{jheppub}
\usepackage{amsmath,amssymb,graphicx,epstopdf,enumerate,amsfonts,booktabs,color,soul}
\usepackage{slashed}
\usepackage{float,placeins}
\usepackage{appendix}
\newcommand{\R}{\mathbb{R}}
\newcommand{\Z}{\mathbb{Z}}
\providecommand{\U}[1]{\protect\rule{.1in}{.1in}}
\affiliation[a]{Department of Electrophysics, National Chiao Tung University, Hsinchu, ROC}
\affiliation[b]{School of physics, University of Chinese Academy of Sciences, Beijing 100049, China}
\affiliation[c]{Kavli Insititute for Theoretical Sciences, University of Chinese Academy of Sciences, Beijing 100049, China}
\emailAdd{sr755332@gmail.com}
\emailAdd{yiyang@mail.nctu.edu.tw}
\emailAdd{phy.pro.phy@gmail.com}
\abstract{The chiral symmetry breaking ($\chi_{SB}$) is one of the most fundamental problems in QCD. In this paper, we calculate quark condensation analytically in a holographic QCD model dual to the Einstein-Maxwell-Dilaton (EMD) system coupled to a probe scalar field. We find that the black hole phase transition in the EMD system seriously affects $\chi_{SB}$. At small chemical potential, $\chi_{SB}$ behaves as a crossover. For large chemical potential $\mu>\mu_c$, $\chi_{SB}$ becomes first order with exactly the same transition temperature as the black hole phase transition by a bypass mechanism. The phase diagram we obtained is qualitatively consistent with the recent results from lattice QCD simulations and NJL models. }
\begin{document}

\title{Analytic Study on Chiral Phase Transition in Holographic QCD}
\author{Meng-Wei Li${^a}$, Yi Yang${^{a}}$, Pei-Hung Yuan${^{b,c}}$}
\maketitle

\setcounter{equation}{0}
\renewcommand{\theequation}{\arabic{section}.\arabic{equation}}

\section{Introduction}
One of the most important and difficult problems in quantum chromodynamics (QCD) is to determine its phase diagram. It is well known that QCD is in the phase of confinement and chiral symmetry breaking ($\chi_{SB}$) at low temperature and small chemical potential (or quark density). On the other hand, at the temperature of the order of $\Lambda_{QCD}\sim 0.2~GeV$, the chiral symmetry is supposed to restore and the color degrees of freedom will be released\footnote{People also conjectured that there is a color-flavor locking or color superconductor phase at low temperature and large chemical potential, but it still remains a mystery.}. It is widely believed that there is a phase transition between the confinement and deconfinement and so does the chiral symmetry breaking and restoring. It is then important to determine the phase boundary between the phases.

\begin{figure}[t!]
\begin{center}
\includegraphics[
height=2in, width=2.8in]
{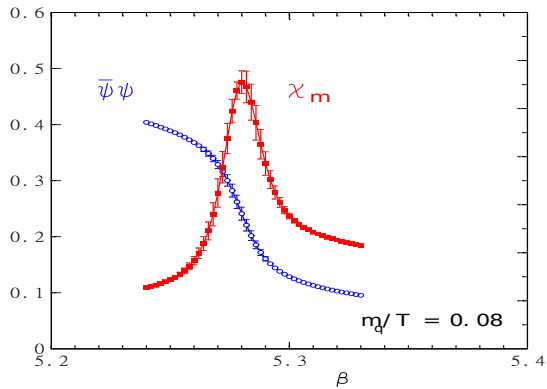}
\end{center}
\caption{ Chiral symmetry restoration of 2-flavour QCD was shown in \cite{0106019}. $\langle \bar{\psi}\psi \rangle$ is the order parameter for chiral symmetry breaking in the chiral limit ($m_q \rightarrow 0$). The corresponding susceptibilities are also demonstrated. }\label{fig_demo}
\end{figure}

However, there are some difficult issues need to be clarified and solved. Firstly, it has been pointed out that the phase boundaries for confinement transition and $\chi_{SB}$ are very close to each other in spite of their quite different origins \cite{0106019,1009.4089}. The mechanism for this overlapped phase boundary phenomenon is not clear yet. Secondly, lattice QCD has showed that the transition is a crossover instead of a first order phase transition around zero chemical potential. On the other hand, people believe that the phase transition should be first order at large enough chemical potential. Therefore, it is crucial to locate the critical end point (CEP) of the phase transition in the phase diagram. Finally, the most important issue is that the phase transition occurs around the strong coupled region so that the conventional perturbation theory in quantum field theory does not work. For a long time, lattice QCD is the only reliable method to attempt this problem. At zero chemical potential, it has been showed that the first order phase transition is demoted to a crossover transition for the physical quark mass. The transition temperatures of the crossover for $\chi_{SB}$ can be determined by calculating quark condensation and identifying the peak of the susceptibility as plotted in Fig.\ref{fig_demo}. Nevertheless, lattice QCD does not work well at finite chemical potential due to the sign problem.

During the last two decades, holographic correspondence has been extensively studied \cite{9711200,9802150,9905111}. One of the important applications of this correspondence is to study the strong coupled quantum systems such as QCD, i.e. the holographic QCD (hQCD). Holographic correspondence offers an ideal frame to study the phase transitions in QCD. A lot of works have been done in this direction, including both top-down \cite{0412141,0507073} and bottom-up \cite{0501022,0501128,0501218,0602229,0804.1096} methods. Because it has to solve a set of highly nonlinear coupled Einstein equations in the bulk space-time, most of the previous works chose to use various numerical methods. Many important results had been obtained, but it is difficult to understand the hidden mechanism without a manifest analytic formula.

In this paper,  inspired by the holographic correspondence, we propose a bottom-up holographic QCD model to analytically study the chiral symmetry transition at finite chemical potential. We use a dynamical soft wall model by considering a 5-dimensional Einstein-Maxwell-Dilaton (EMD) system whose fields content is arranged to describe the dynamics of $\chi_{SB}$  in QCD. A family of analytic black hole solutions of this model has been obtained previously in \cite{1201.0820,1301.0385,1406.1865,1506.05930,1703.09184,1705.07587,1812.09676,2004.01965}. By studying the thermodynamics of this system, we investigate the phase transition between black holes in the 5-dimensional bulk space-time, as well as the equations of states, including thermal entropy, specific heat, speed of sound, and the trace anomaly.

Confinement-deconfinement phase transition has been analytically studied in EMD system by adding open strings in the bulk space-time in \cite{1506.05930,1703.09184}. Hence, we will focus on $\chi_{SB}$ in this work. To study $\chi_{SB}$, we add a composite scalar as a probe field into the background. The expansion of the scalar field at the boundary gives the temperature dependent quark condensation which is the order parameter of $\chi_{SB}$. An analytic solution of the scalar field can be obtained by a matching method. The asymptotic solutions near both the boundary and the horizon are obtained by solving the composite scalar field order by order. Smoothly matching the two asymptotic solutions at an intermediate matching point gives an analytic solution of the scalar. The matching method has been use to study $\chi_{SB}$ at zero chemical potential \cite{1112.4402,1206.2824,1303.6929,1411.5332,1511.02721,1512.04062,1512.06439,1610.09814,1810.07019}. However, the solution obtained from the matching method is an approximate one. We will justify the validity of the solution by comparing our results with the numeric ones. We find that the $\chi_{SB}$ transition is closely related to the phase transition between black holes in the bulk space-time.

There is an important issue we would like to mention: the mass dependent of the phase transition behavior in QCD. Lattice simulation \cite{1009.4089} has shown that, at $\mu=0$, it is the first order phase transition in both the chiral limit $m_{q}\rightarrow0$ and the quench limit $m_{q}\rightarrow\infty$, and becomes a crossover in between. There exists two boundaries, on which the phase transition becomes second order, between the crossover and the two first order phase transition regions. It has been shown that the crossover region shrinks as $\mu$ growing, so that we expect a critical point at a finite $\mu$ where the crossover changes to the first order phase transition. So the question is: where is the location of the boundaries at both zero and finite $\mu$ in massive quark system? To our knowledge, there is no bottom-up holographic QCD model seriously discussing this issue so far. In this work, we will not discuss more about the mass dependent issue and only focus on the case with the physical quark mass $m_{q}\sim3MeV$. We leave this mass dependent issue in QCD phase transition in the future work.

This paper is organized as the following. In section II, we review the EMD system and the analytic black hole solutions. The thermodynamic properties of the bulk space-time are studied to obtain the phase structure of the black hole background. We then add a probe scalar into the background and solve the scalar field analytically by the matching method in section III. Furthermore, we investigate the phase diagram of $\chi_{SB}$ by combining the $\chi_{SB}$ transition and the back hole phase transition in the background. Section IV contains our conclusion.

\section{EMD System}
EMD system is one of the most fundamental framework to establish the dual gravitational theory of the strongly coupled quantum field theory. By using EMD system to construct hQCD models was advocated from \cite{1012.1864, 1108.2029}, and has been improved afterward in many literature. In this work, we will analytically investigate the QCD phase structure by using a hQCD model in \cite{1301.0385,1406.1865,1506.05930,1703.09184,1705.07587,1812.09676}. 
 
\subsection{EMD Background}
We consider a 5-dimensional EMS system with probe matter fields. \cite{1301.0385,1406.1865,1506.05930,1703.09184,1705.07587,1812.09676,2004.01965}. The system can be described by an action with two parts, the background sector $S_{B}$ and the matter sector $S_{m}$,
\begin{equation}
S=S_{B}+S_{m}. \label{eq_S}
\end{equation}
In the string frame, labeled by a sup-index $s$, the background sector $S_{B}$ includes a gravity field $g^s_{\mu\nu}$, a Maxwell field $A_{\mu}$ and a neutral scalar field $\phi^s$ with action,
\begin{equation}
S_{B} = \frac{1}{16\pi G_{5}} \int d^{5}x\sqrt{-g^s}e^{-2\phi^s}
	\left[{R^s-\frac{f^s_B(\phi^s)}{4}{F}^{2}}
			+4\partial_{\mu}\phi^s\partial^{\mu}\phi^s
			-V^s(\phi^s)  \right], \label{eq_SB_sf}
\end{equation}
where ${G}_{5}$ is the 5-dimensional Newtonian constant and ${{F}}_{\mu\nu} = \partial_{\mu}{A}_{\nu}-\partial_{\nu}{A}_{\mu}$ is the gauge field strength corresponding to the Maxwell field. The leading term of the gauge field expansion near the AdS boundary is associated to the chemical potential $\mu$ in hQCD. The function $f^s(\phi^s)$ is the gauge kinetic function associated to Maxwell field and $V^s(\phi^s)$ is the potential of the scalar field.

The action of the matter sector includes two parts $S_{m}=S_V+S_\chi$ with
\begin{eqnarray}
S_{V} & =& -\frac{1}{16\pi G_{5}}\int d^{5}x\sqrt{-g^s}e^{-\phi^s}
		\left[{\frac{f^s\left(\phi^s\right)}{4}} 
		\left(F_{V}^{2}+F_{\tilde{V}}^2\right)\right], \label{SV}\\
S_\chi &=&-\int{}^{}d^5x\sqrt{-g^s} e^{-\phi _s} Tr\{\nabla_{M}X^{\dagger}\nabla^{M}X+m_{\chi}^2X^{\dagger}X\}, \label{SX}
\end{eqnarray}
where the massless gauge fields $A_{\mu}^{V}$ and $A_{\mu}^{\tilde{V}}$ describing the degrees of freedom of vector mesons and pseudovector mesons on the 4-dimensional boundary, and $S_\chi$ is the action of the composite scalar $X$ describing the quark condensation $\langle\bar\psi \psi\rangle$ in the vacuum with $m_\chi$ being the mass of the 5-dimensional scalar. We treat the matter fields as probe and ignore their back-reaction to the background.

We have described the EMS system in string frame. However, for the background part, it is more convenient to solve the equations of motion and study the thermodynamical properties of hQCD in Einstein frame, which can be obtained by the following Weyl transformations,
\begin{equation} \label{eq_weyl trans}
\phi^{s}=\sqrt{\frac{3}{8}}\phi ,~
g^s_{\mu\nu}=g_{\mu\nu} {e}^{\sqrt{\frac{2}{3}}\phi} ,~
f^{s}\left(\phi^{s}\right)=f\left(\phi\right) {e}^{\sqrt{\frac{2}{3}}\phi} ,~
V^{s}\left(\phi^{s}\right)={e}^{-\sqrt{\frac{2}{3}}\phi}V\left(\phi\right).
\end{equation}
The action of the background Eq. (\ref{eq_SB_sf}) becomes
\begin{eqnarray}
S_{B} &=& \frac{1}{16\pi G_{5}} \int d^{5}x\sqrt{-g}
	\left[{R-\frac{f\left(\phi\right)}{4}F^{2}}
			-\frac{1}{2}\partial_{\mu}\phi \partial^{\mu}\phi
			-V\left(\phi\right) \right]. \label{eq_SB_Ef}
\end{eqnarray}
The equations of motion for the background can be derived as,
\begin{eqnarray}
\nabla^{2}\phi &=& \frac{\partial V}{\partial\phi}+\frac{F^2}{4}\frac{\partial f}{\partial\phi}, \label{eq_eom_phi}\\
\nabla_{\mu}\left[ f(\phi)F^{\mu\nu} \right] &=&0, \label{eq_eom_A}\\
R_{\mu\nu}-\frac{1}{2} g_{\mu\nu}R &=& \frac{f(\phi)}{2} \left( F_{\mu\rho}F_{\nu}^{~\rho}-\frac{1}{4}g_{\mu\nu}F^{2}\right) +\frac{1}{2}\left[\partial_{\mu}\phi\partial_{\nu}\phi-\frac{1}{2}g_{\mu\nu}\left(\partial\phi\right)^{2}-g_{\mu\nu}V(\phi)\right] \label{eq_eom_g}.
\end{eqnarray}
Since we are going to study the phase transitions in QCD at finite temperature and chemical potential, without loss of generality, we consider the following ansatz for the metric of the bulk space-time in Einstein frame,
\begin{eqnarray}
ds^{2} &=& \frac{e^{2A\left(z\right)}}{z^{2}}
			\left[-g(z)dt^{2} +d\vec{x}^{2} +\frac{dz^{2}}{g(z)}\right],\label{eq_metric}\\
\phi &=& \phi\left(z\right),~ A_{\mu}=\left(A_{t}\left(z\right),\vec{0},0\right), \label{eq_ansatz}
\end{eqnarray}
where $z = 0$ corresponds to the conformal boundary of the 5-dimensional space-time and $g(z)$ stands for the blackening factor. Here we have set the radial of $AdS_5$ to be unit by a scale invariant of the system.

Plugging the ansatz Eq. (\ref{eq_metric}) and Eq. (\ref{eq_ansatz}) into Eqs. (\ref{eq_eom_phi}-\ref{eq_eom_g}), the equations of motion become
\begin{eqnarray}
\phi^{\prime\prime}+\left(\frac{g^{\prime}}{g}+3A^{\prime}-\dfrac{3}{z}\right) \phi^{\prime}+\left( \frac{z^{2}e^{-2A}A_{t}^{\prime2}f_{\phi}}{2g}-\frac{e^{2A}V_{\phi}}{z^{2}g}\right)   &=&0,\label{eom_phi}\\
A_{t}^{\prime\prime}+\left(  \frac{f^{\prime}}{f}+A^{\prime}-\dfrac{1}{z}\right)  A_{t}^{\prime}  &=&0,\label{eom_At}\\
A^{\prime\prime}-A^{\prime2}+\dfrac{2}{z}A^{\prime}+\dfrac{\phi^{\prime2}}{6}&=& 0,\label{eom_A}\\
g^{\prime\prime}+ 3g'\left(  A^{\prime}-\dfrac{1}{z}\right) - \frac{fz^2 A_t'^2}{e^{2A}} &=& 0,\label{eom_g}\\
A''+3A'^2-\frac{2}{z}A'+\left( A'-\frac{1}{z} \right) \left( \frac{3g'}{2g}-\dfrac{4}{z}\right) +\dfrac{g''}{6g}+\frac{e^{2A}V}{3z^{2}g}  &=& 0.
\label{eom_V}
\end{eqnarray}

Before we start to solve the equations of motion, it is worth to verify the null energy condition (NEC) to guarantee the stability of the gravitational model. The NEC can be expressed as
\begin{equation}
T_{\mu \nu}N^{\mu}N^{\nu} \geq 0, \label{NEC}
\end{equation}
where $T^{\mu\nu}$ is the energy-momentum tensor of the background matter fields. The null vector $N^\mu$ satisfies the condition $g_{\mu\nu}N^{\mu}N^{\nu}=0$ and could be chosen as 
\begin{equation}
   N^{\mu}=\frac{1}{\sqrt{g\left(  z\right)  }}N^t
          +\frac{\sin\theta}{\sqrt{3}} N^{\vec{x}}
          + \cos\theta\sqrt{g\left(z\right)} N^{z},
\end{equation}
for arbitrary parameter $\theta$. Then the NEC Eq. (\ref{NEC}) becomes
\begin{equation}
\frac{1}{2}\left( f \frac{A_t'^2 z^2 \sin^2\theta}{e^{2A}}  +g \phi'^2 \cos^2\theta\right) \geq 0, \label{NEC constraint}
\end{equation}
which, together with the above equations of motion, requests that the gauge kinematic function is non-negative and the scalar field $\phi$ is real, i.e. $f \geq 0$ and $\phi'^2 \geq 0$. Importantly, the condition $\phi'^2 \geq 0$ constraints the functional form of the warped function $A(z)$ in Eq. (\ref{eq_metric}).

To solve the above equations of motion, proper boundary conditions should be imposed. At the boundary $z=0$, we impose that the metric in the string frame is asymptotic to $AdS_5$, that leads
\begin{equation}
 		A(0)+\sqrt{\frac{1}{6}}\phi(0)=0,~ g(0)=1. \label{bdy_0}
		\end{equation}
At the horizon $z=z_H$, we require the regularity of the black hole solution,
\begin{equation}
 		A_t(z_H)=g(z_H)=0. \label{bdy_zh}
		\end{equation}
By using the potential reconstruction method, the equations of motion Eqs. (\ref{eom_phi}-\ref{eom_g}) can be solved analytically as,
\begin{eqnarray}
\phi\left( z \right) &=&\int_0^z dy\sqrt{-6\left( A^{\prime\prime}
-A^{\prime2}+\dfrac{2}{z}A^{\prime}\right)  },\label{phip-A}\\
A_{t}\left(  z\right)   &=& \mu \left(1-\frac{\int_0^{z}\frac{y}{fe^{A}}dy}{\int_0^{z_H}\frac{y}{fe^{A}}dy}\right)=\mu-\rho z^2 +\cdots,\label{At-A}\\
g\left(  {z}\right)   &=&1-\frac{\int_{0}^{z} \frac{y^3}{e^{3A}} dy}{\int_{0}^{z_{H}%
} \frac{y^3}{e^{3A}} dy}+\dfrac{\mu^{2}\left\vert
\begin{array}
[c]{cc}%
\int_{0}^{z_{H}} \frac{y^3}{e^{3A}} dy & \int_{0}^{z_{H}}\frac{y^3}{e^{3A}}dy\int_{0}^{y}%
\frac{x}{fe^{A}}dx\\
\int_{z_{H}}^{z}\frac{y^3}{e^{3A}}dy & \int_{z_{H}}^{z}\frac{y^3}{e^{3A}}dy\int_{0}^{y}%
\frac{x}{fe^{A}}dx
\end{array}
\right\vert }{ \left( \int_{0}^{z_{H}} \frac{y^3}{e^{3A}} dz \right) \left( \int_{0}^{z_{H}} \frac{z}{fe^{A}}dz \right)^{2}}, \\
V(z) &=&-\frac{3g z^2}{e^{2A}}
        \left[
            \left( A''+3A'^2-\frac{2}{z}A' \right)
            +\left( A'-\frac{1}{z} \right) \left( \frac{3g'}{2g}-\dfrac{4}{z}\right) +\dfrac{g''}{6g}
        \right]. \label{V-A}
\end{eqnarray}
where $\mu$ is the chemical potential according to the holographic dictionary of the AdS/CFT correspondence, and $\rho$ stands for the baryon density that relates to the chemical potential as,
\begin{equation}
\rho=\dfrac{\mu}{2 \int_{0}^{z_{H}}\dfrac{y}{fe^A}dy}. \label{rho}%
\end{equation}
Eqs. (\ref{phip-A}-\ref{V-A}) represent a family of black hole solutions, the different choice of the functions $A$ and $f$ corresponds to the different solution.

\subsection{Meson Mass Spectrum}
One of the crucial properties of the soft-wall hQCD models is that the vector meson spectrum satisfies the linear Regge trajectories at zero temperature. This issue was first addressed in \cite{0602229} by holographic correspondence.

To study this issue in our model, we consider the probe vector field $V$ in the bulk space-time. The equation of motion for the vector field can be obtained by varying the action Eq. (\ref{SV}),
\begin{equation}
\nabla_{\mu}\left[e^{\sqrt{\frac{3}{8}\phi}} f\left( \phi\right) F_{V}^{\mu\nu}\right]  =0,
\end{equation}
in Einstein frame.

By fixing the gauge $V_{z}=0$, the equation of motion of the transverse part of the vector field $V_{\mu}$ $\left( \partial^{\mu}V_{\mu}=0\right) $ in the background Eq. (\ref{eq_metric}) reduces to a Schr\"{o}dinger equation,
\begin{equation}
-\psi_{i}^{\prime\prime}+U\left( z\right) \psi_{i}=\left( \dfrac{\omega
^{2}}{g^{2}}-\dfrac{p^{2}}{g}\right) \psi_{i}, \label{eqv}%
\end{equation}
where we have performed the Fourier transformation for the vector field $V_{i}$,
\begin{equation}
V_{i}\left( x,z\right) =\int\dfrac{d^{4}k}{\left( 2\pi\right)^{4}%
}e^{ik\cdot x}v_{i}\left( z\right) , \label{V-v}%
\end{equation}
and made a transformation,
\begin{equation}
v_{i}=\left( \dfrac{z}{e^{\sqrt{\frac{3}{8}\phi}} e^{A}fg}\right)^{1/2}\psi_{i}\equiv X\psi_{i}.
\end{equation}
The potential function $U$ in the Schr\"{o}dinger equation Eq. (\ref{eqv}) is defined as
\begin{equation}
U\left( z\right) =\dfrac{2X^{\prime2}}{X^{2}}-\dfrac{X^{\prime\prime}}{X}.
\end{equation}
At zero temperature, $g\rightarrow1$, we expect that the discrete spectrum of the vector mesons obeys the linear Regge trajectories Eq. (\ref{eqv}) in the zero temperature limit reduces to the following form,
\begin{equation}
-\psi_{i}^{\prime\prime}+U\left(  z\right)  \psi_{i}=m^{2}\psi_{i},
\label{Schrodinger}%
\end{equation}
where $m^{2}=-k^{2}=\omega^{2}-p^{2}$. To produce the discrete mass spectrum which satisfied with the linear Regge trajectories, the potential $U\left( z \right)$ should be in certain forms. A simple choice for the gauge kinetic function is
\begin{equation}
f\left( z\right) = e^{-\sqrt{\frac{3}{8}\phi}-cz^{2}-A\left(  z\right)  },
\end{equation}
which leads the potential to be
\begin{equation}
U\left( z\right) = -\dfrac{3}{4z^{2}}-c^{2}z^{2}. \label{potential}
\end{equation}
The Schr\"{o}dinger Eq. (\ref{Schrodinger}) with the potential Eq. (\ref{potential}) has the discrete eigenvalues,
\begin{equation}
m_{n}^{2}=4cn, \label{mass}%
\end{equation}
which is the well known linear Regge trajectories \cite{0507246}, where $n$ stands for the energy level. By fitting the mass spectrum with $\rho$ meson tower, we fixed the parameter $c=0.227 ~GeV$, \cite{1406.1865}.

Once we fixed the gauge kinetic function $f(z)$, the solutions Eqs. (\ref{phip-A}-\ref{V-A}) represent a family of solutions for the black hole background depending on the choice of the warped factor $A\left(z\right)$ which satisfies the boundary condition in Eq. (\ref{bdy_0}). Following \cite{1703.09184,1705.07587,1812.09676,2004.01965}, we choose the warped factor to be 
\begin{equation}
A \left( z \right)= -a \ln (b z^2+1) \label{ansatz_A}
\end{equation}
with the parameters $a=4.046$ and $b=0.01613$, which are fixed by comparing the confinement phase transition temperature at zero chemical potential with the lattice QCD result $T \simeq 157 MeV$ in \cite{1701.04325}.

\subsection{Phase Transition in the Background} \label{BH PTs}
\begin{figure}[t!]
\begin{center}
\includegraphics[
height=2.in, width=2.8in]
{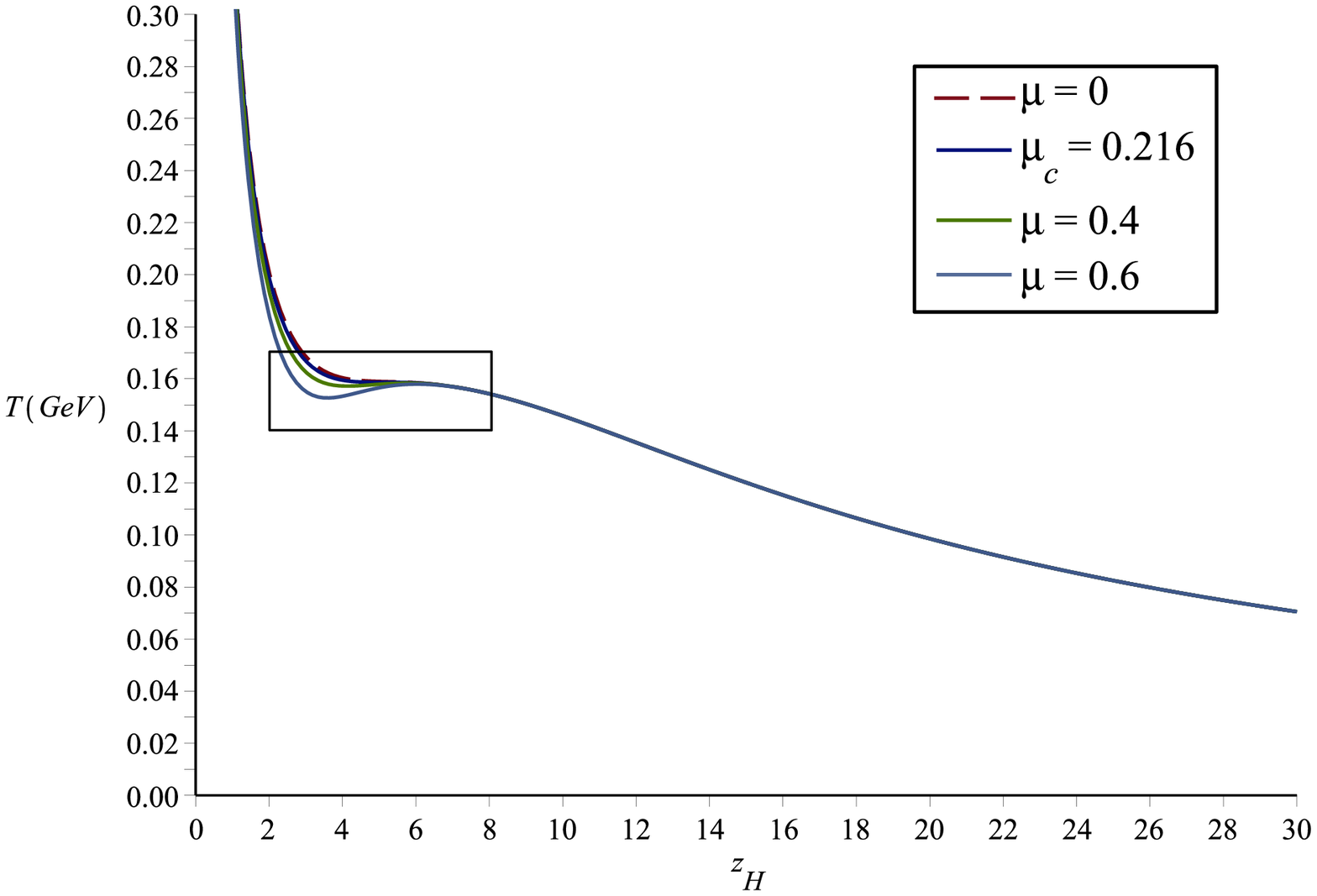}
\includegraphics[
height=2.in, width=2.8in]
{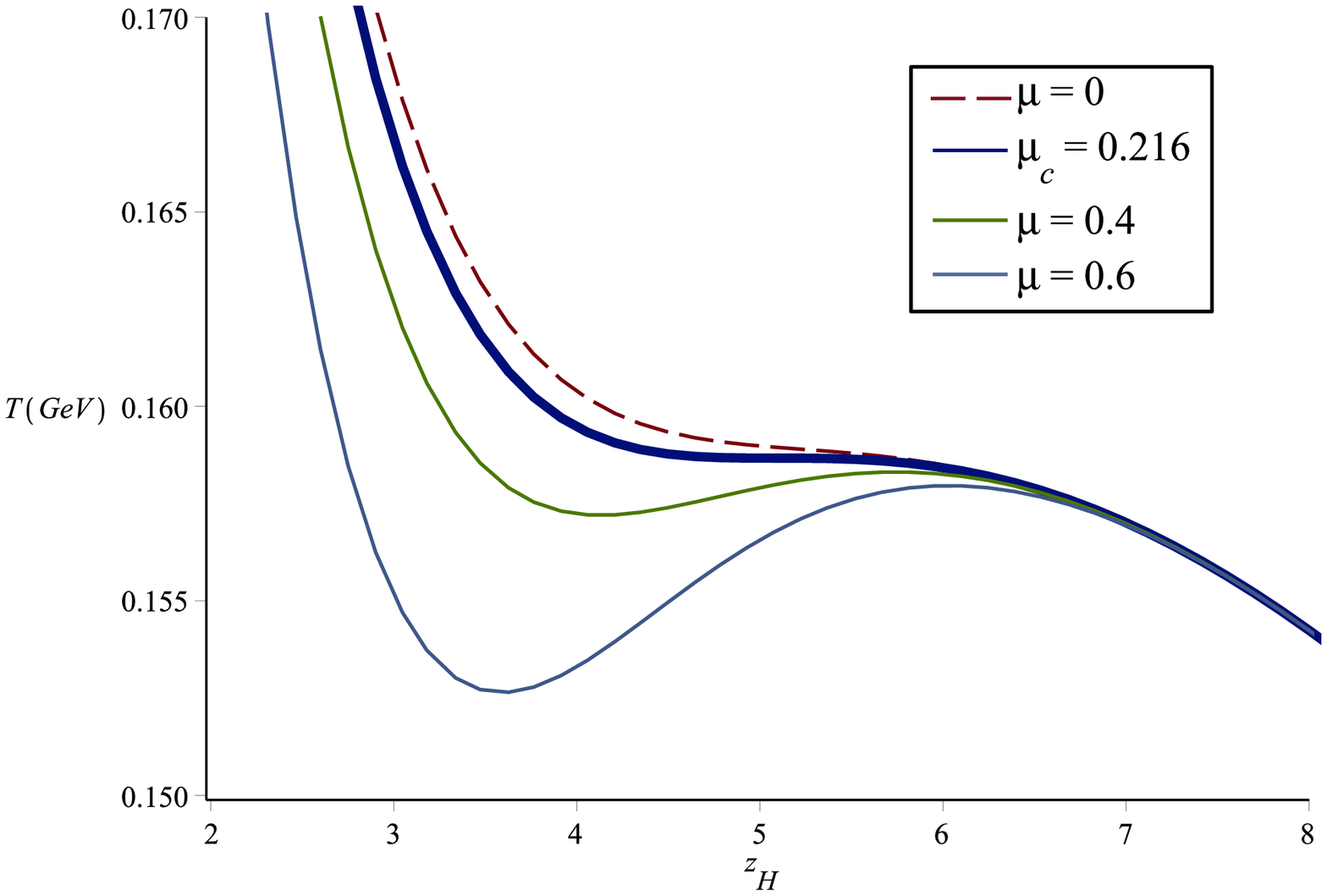}
\vskip -0.05cm \hskip 1 cm (a) \hskip 7 cm (b)
\end{center}
\caption{ (a) The black hole temperature vs. horizon at different chemical potentials. We enlarge the phase transition region in (b) to display the detail structure. For $\mu>mu_c$, the temperature becomes multi-valued that implies a first order phase transition. The transition temperature can be determined by calculating the black hole free energy. } \label{fig_T}
\end{figure}

In this section, we study the phase structure of the black hole background in Eqs. (\ref{phip-A}-\ref{V-A}) obtained in the last section. The Hawking temperature can be calculated as,
\begin{eqnarray}
T&=&\dfrac{z_{H}^{3}e^{-3A\left( z_{H} \right) }} {4\pi \int_{0}^{z_{H}} \frac{y^{3}}{e^{3A}}dy}
    \left[ 1-2c\mu^{2}
    \frac{ e^{cz_{H}^{2}}\int_{0}^{z_{H}}\frac{y^{3}}{e^{3A}}dy-\int_{0}^{z_{H}}\frac{y^{3}}{e^{3A-cy^{2}}}dy  } {\left( 1-e^{cz_{H}^{2}} \right)^{2}}
    \right]  .
\end{eqnarray}
The temperature vs. horizon at different chemical potentials are plotted in Fig. \ref{fig_T}. At small chemical potential, $0 \leq \mu \leq \mu_c$, where $\mu_c=0.216~GeV$ labels the critical chemical potential, the temperature is a monotonously decreasing function of horizon. While at large chemical potential, $\mu > \mu_c$, the temperature becomes multi-valued implying a phase transition between black holes with different sizes.

To determine the phase transition temperatures at different chemical potentials, it is necessary to consider the free energy. In grand canonical ensemble, the free energy is defined by the first law of thermodynamics,
\begin{equation}
dF = -sdT-\rho d\mu.
\end{equation}For a fixed chemical potential $\mu$, the free energy can be evaluated by the following integral,
\begin{equation}
F = -\int sdT= \int_{zH}^\infty s(z_H)T'(z_H)dz_H, \label{eq_int_F}
\end{equation}
where 
\begin{eqnarray}
s = \frac{e^{3A(z_H)}}{4z_H^{3}}, \label{entropy}
\end{eqnarray}
is the black hole entropy and we have normalized the free energy to vanish at $z_H \to \infty$, i.e. $T=0$.

\begin{figure}[t!]
\begin{center}
\includegraphics[
height=2in, width=3.2in]
{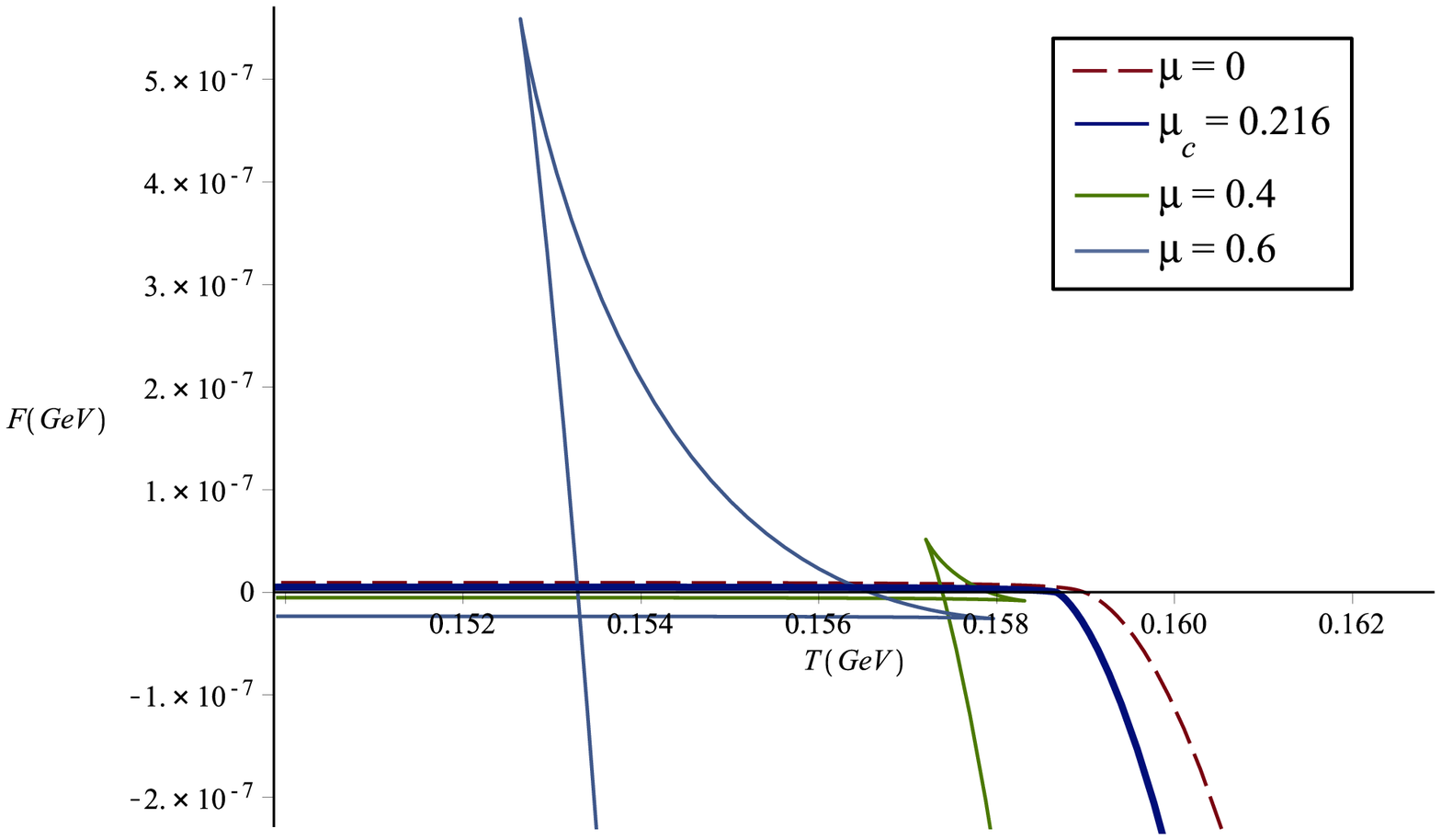}
\includegraphics[
height=2in, width=2.8in]
{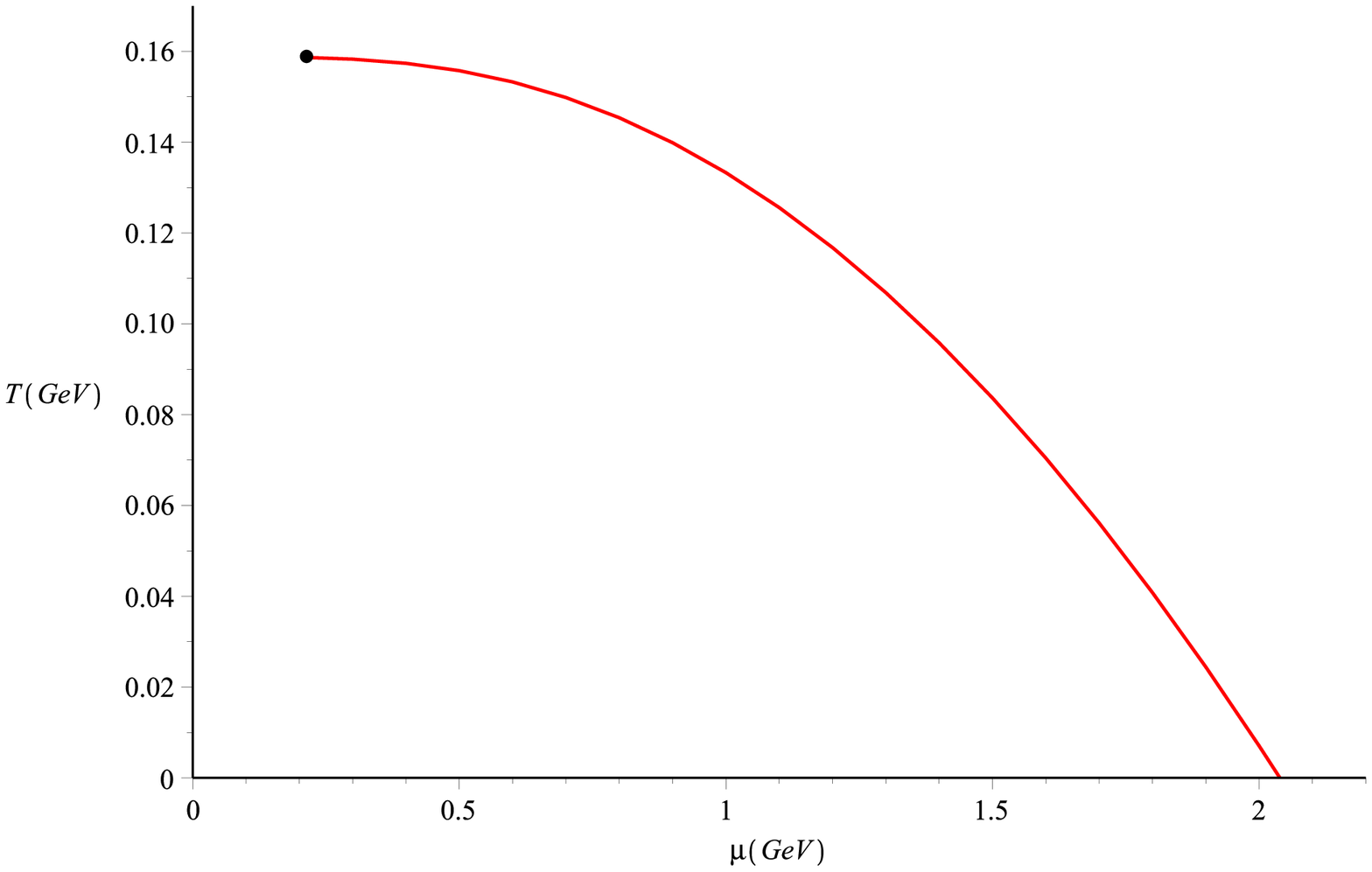}
\vskip -0.05cm \hskip 1 cm (a) \hskip 7 cm (b)
\end{center}
\caption{ (a) The free energy vs. temperature at different chemical potentials. (b) The phase diagram in $T-\mu$ plane. The red line represents the first order phase transition and the black dot labels the critical endpoint at $(\mu_c,T_c) \simeq (0.216,0.159)$. The zero temperature phase transition is located at $\mu_{T=0}\simeq2.039$. } \label{Tmu_BH}
\end{figure}

Eq. (\ref{eq_int_F}) can be integrate numerically to obtain the free energy. The free energy vs. temperature at different chemical potentials are plotted in Fig. \ref{Tmu_BH}(a). For $\mu>\mu_c$, the free energy behaves as the swallow-tiled shape. The intersection of the free energy curve gives the transition temperature of the phase transition between two black holes with different sizes. As the chemical potential decreases, the size of the swallow-tiled shrinks. At the critical chemical potential $\mu=\mu_c$,  the swallow-tiled reduces a singular point, and finally disappears for $\mu<\mu_c$. The behavior of the free energy exhibits that the system undergoes a first order phase transition at the large chemical potential $\mu>\mu_c$ that ends at a CEP at $(\mu_c,T_c) \simeq (0.216,0.159)$ where the phase transition becomes second order. For $\mu<\mu_c$, the phase transition reduces to a crossover. The phase diagram of the black holes phase transition in the bulk background is plotted in Fig. \ref{Tmu_BH}(b).

\subsection{Equations of States}
\begin{figure}[t!]
\begin{center}
\includegraphics[
height=2.in, width=2.8in]
{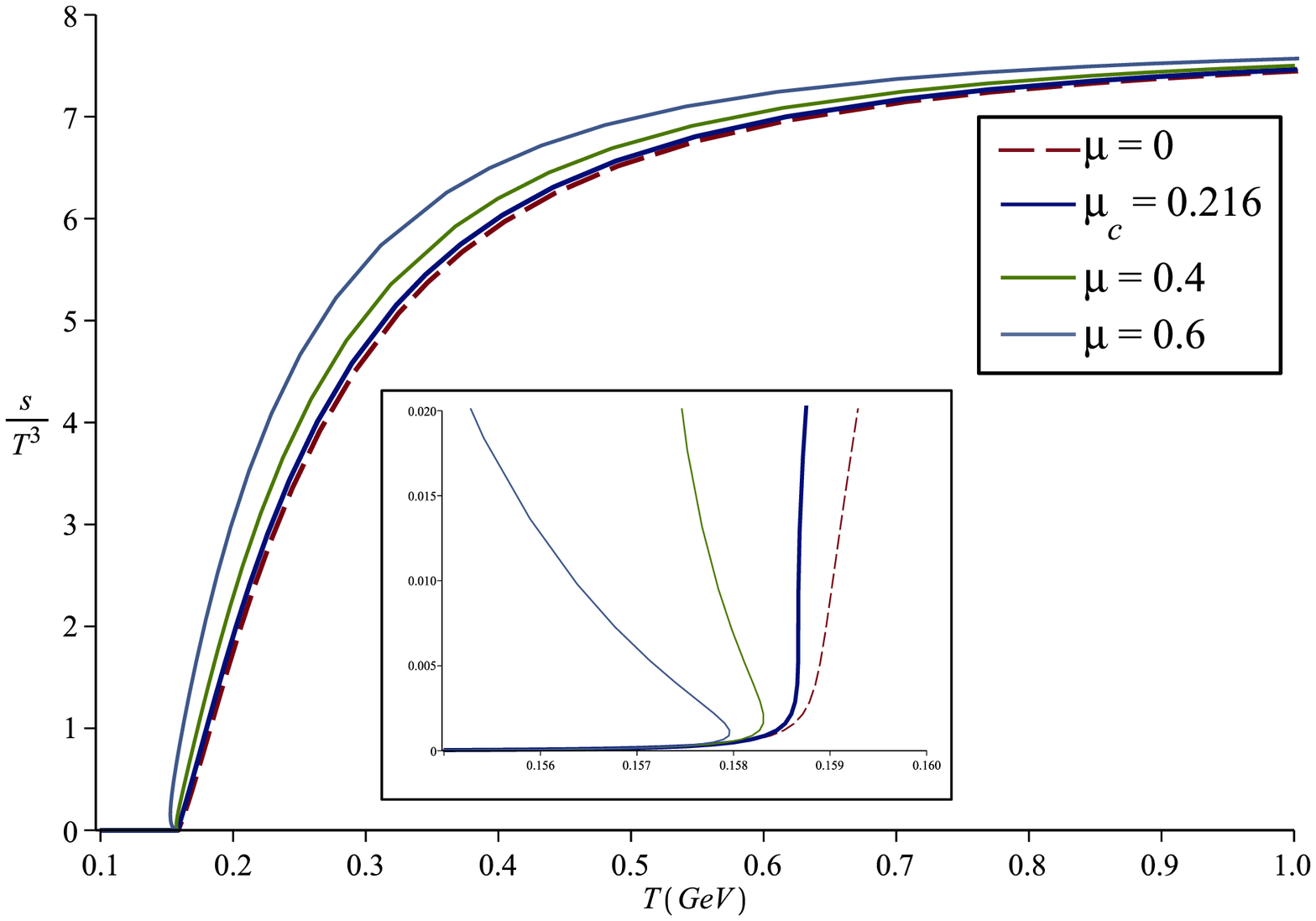}
\includegraphics[
height=2.in, width=2.8in]
{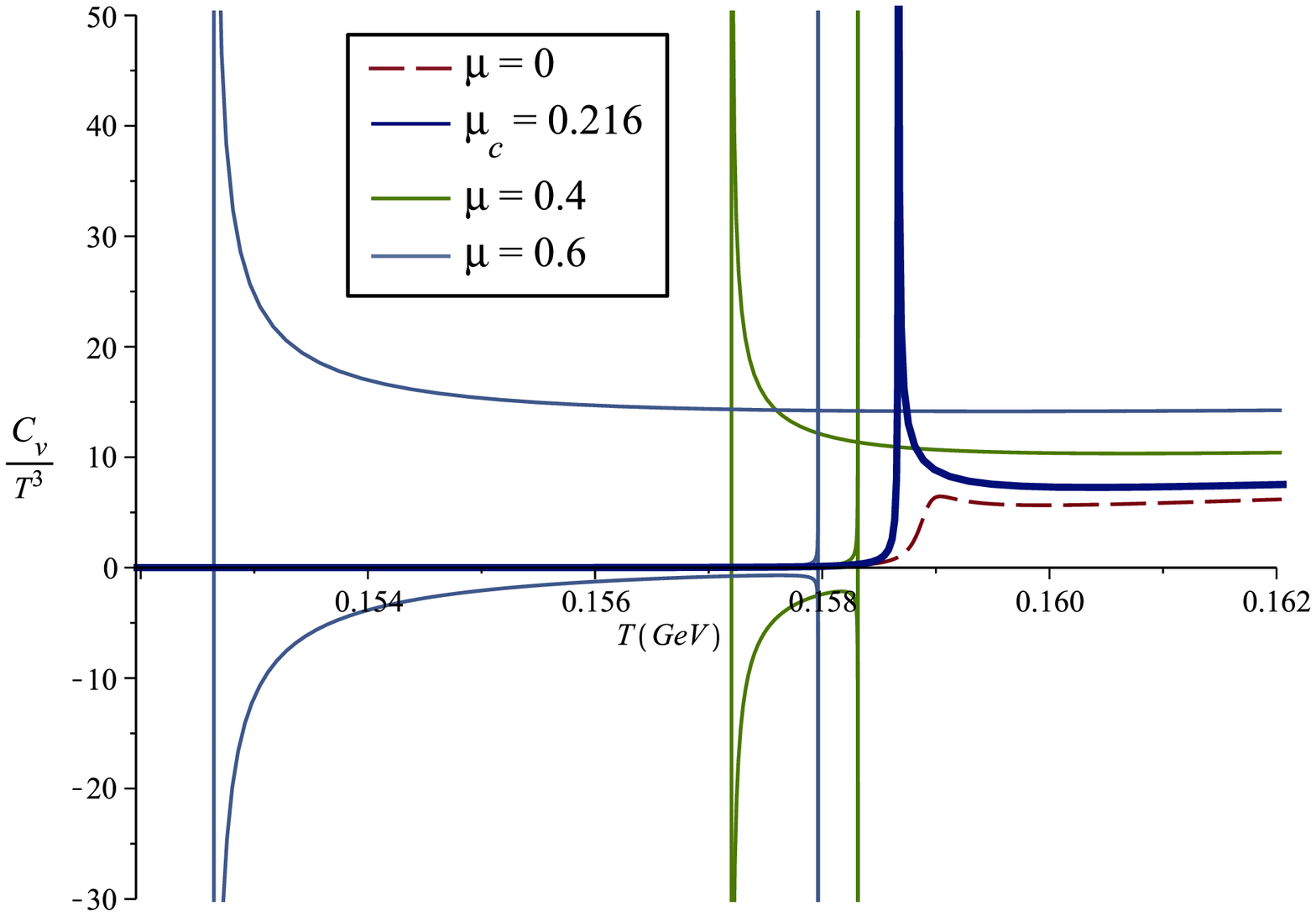}
\vskip -0.05cm \hskip 1 cm (a) \hskip 7 cm (b) \\
\vskip 0.7cm
\includegraphics[
height=2.in, width=2.8in]
{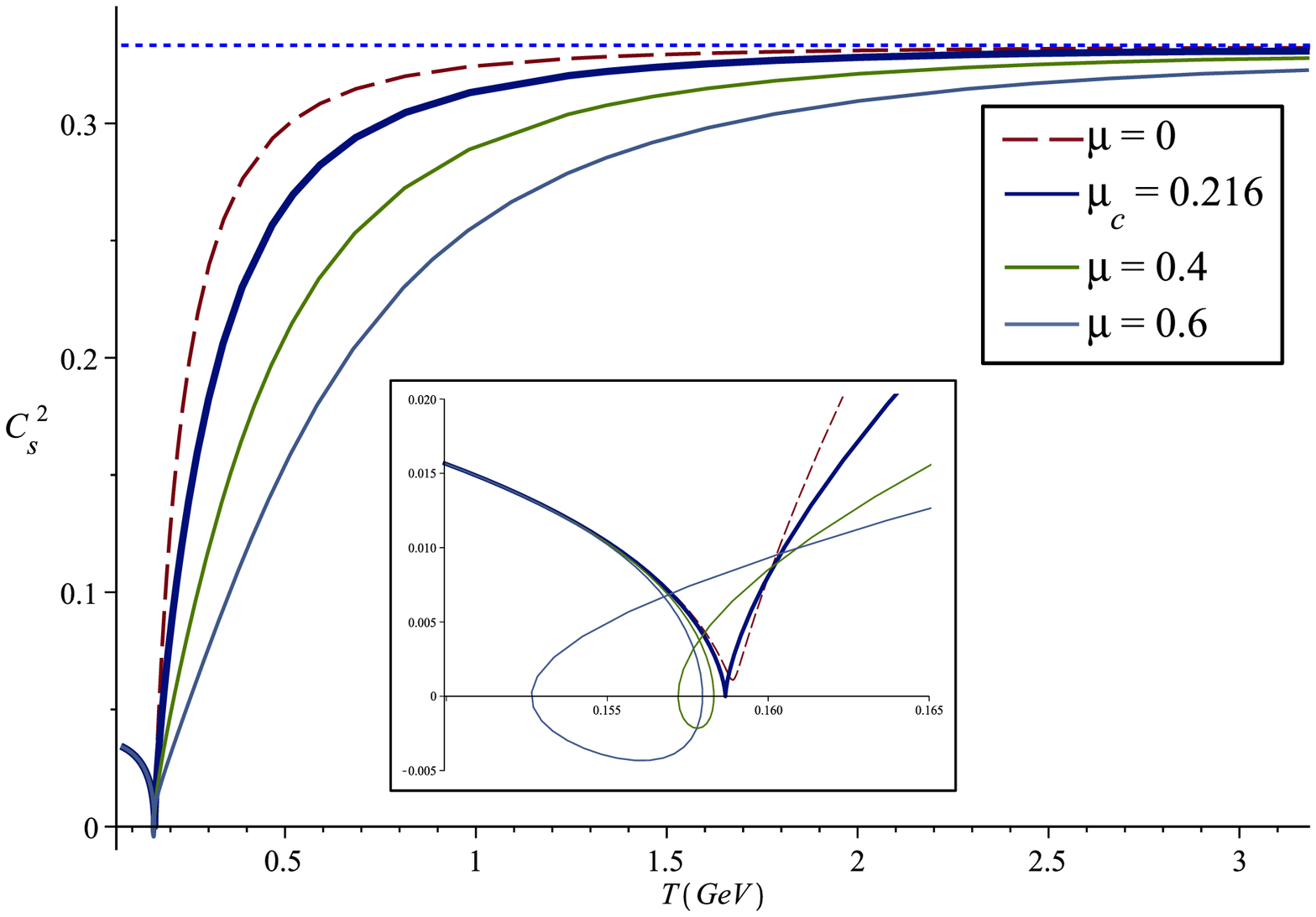}
\includegraphics[
height=2.in, width=2.8in]
{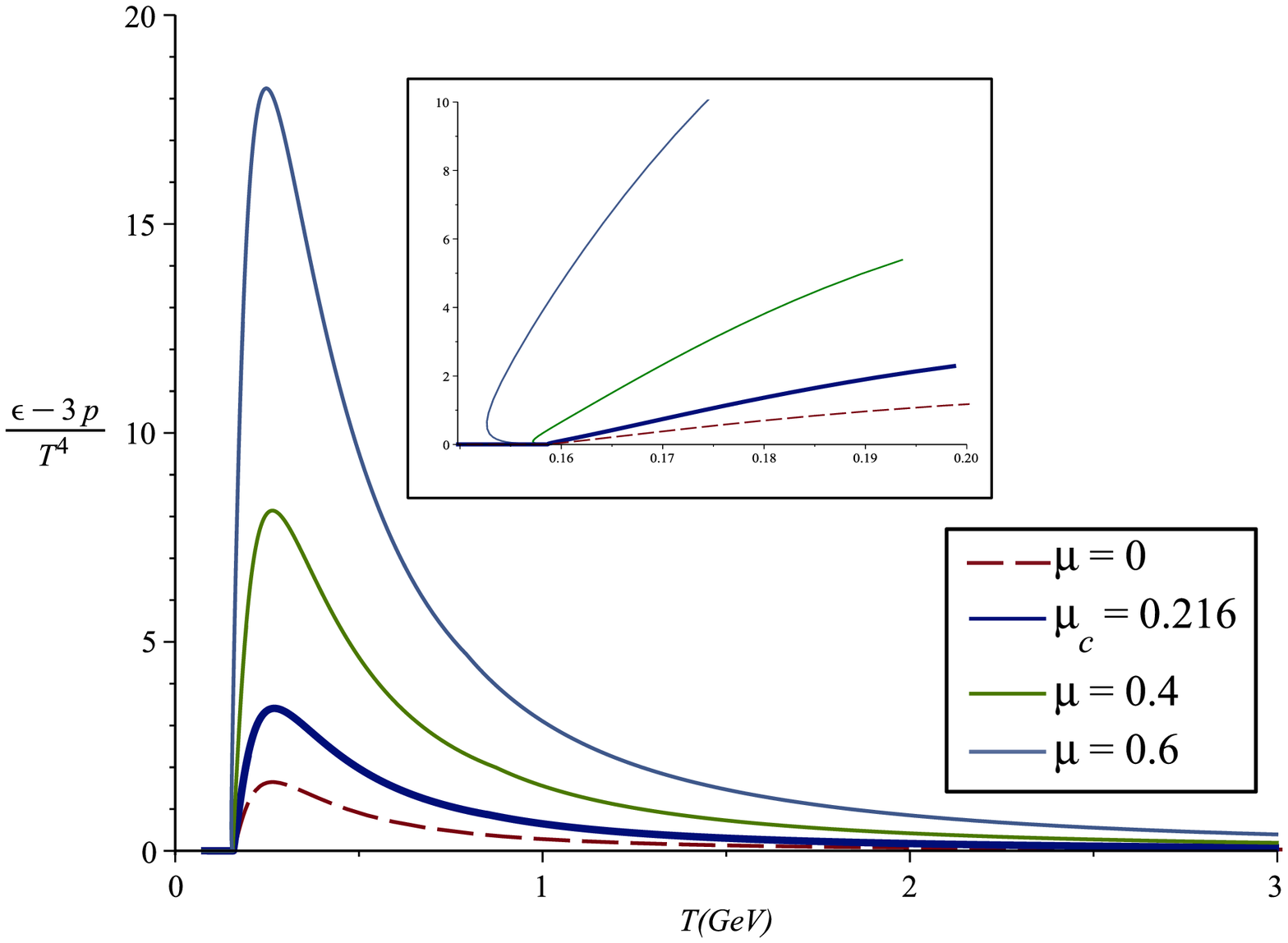}
\vskip -0.05cm \hskip 1 cm (c) \hskip 7 cm (d)
\end{center}
mM\caption{ The entropy, specific heat, speed of sound and trace anomally vs. temperature at different chemical potentials are showed in (a, b, c, d). The phase transition region is enlarged. } \label{fig_EOS}
\end{figure}
The entropy density $s$ is an important thermodynamic quantity which has been defined in Eq. (\ref{entropy}). The dimensionless quantity $s/T^3$ vs. temperature is plotted in Fig. \ref{fig_EOS}(a). In the high temperature limit $T\to \infty$, $s/T^3$ approaches the hadronic freeze-out conditions \cite{1603.03847,1807.08105}, $s/T^3\to \pi^3/4$, which is independent of the chemical potential. On the other hand, $s/T^3$ around the phase transition temperature $T\sim 0.15~GeV$ is multi-valued as is enlarged in Fig. \ref{fig_EOS}(a). Under the phase transition, the entropy density jumps from almost zero to a finite value.

Another important thermodynamic quantity is the specific heat $C_V$, which not only signifies the stability of the black hole solutions  but also implies the phase transition \cite{2006.03494}. The specific heat $C_V$ and can be defined as
\begin{equation}
C_{V}=T\frac{\partial s}{\partial T}.
\end{equation}
The dimensionless quantity $C_V/T^3$ vs. temperature is plotted in Fig. \ref{fig_EOS}(b). The negative branch of the specific heat corresponds to the thermodynamic instability. For $\mu < \mu_c$, the specific heat is always positive indicating that the black hole solutions are always thermodynamically stable. As the chemical potential increasing, when $\mu > \mu_c$, the specific heat becomes multi-valued and the negative branch of the specific heat emerges indicating that the black hole is thermodynamically unstable in this branch. This negative branch corresponds to the negative-slope segment between the local maximum and minimum temperatures as showed in Fig. (\ref{fig_T}).

Furthermore, the speed of sound plays a characteristic role of the thermodynamic quantity in QCD and quark-gluon plasma states. For non-zero chemical potential conditions, the speed of sound can be calculated by \cite{1705.07587,1707.00872}
\begin{equation}
c_s^2=\frac{s}{T\left(\frac{\partial s}{\partial T}\right)_\mu+\mu \left(\frac{\partial \rho}{\partial T}\right)_\mu}.
\end{equation} 
The square of speed of sound $c_s^2$ vs. temperature is plotted in Fig. \ref{fig_EOS}(c). In the high temperature limit, $c_s^2$ approaches the conformal limit $1/3$ as expected.  For $\mu<\mu_c$, the speed of sound is a smooth curve with a rapid turning point around the crossover transition temperature. The turning point sharpens to a tip touching zero at  $\mu=\mu_c$. For $\mu>\mu_c$, The speed of sound becomes multi-valued with the intersect point locating the phase transition temperature. In addition, a negative branch emerges for  $c_s^2$ induces the imaginary of the speed of sound that implies the Gregory-Laflamme dynamical instability \cite{9301052,9404071}. In our system, this dynamical instability is equivalent to the thermodynamic instability from the specific heat as the Gubser-Mitra conjecture \cite{0009126,0011127,0104071}.

Finally, the trace anomaly $\epsilon-3p$ is the sign of the system deviated from the conformality, where $p=-F$ and $\varepsilon=F+Ts+\mu\rho$. The dimensionless quantity $(\epsilon-3p)/T^4$ vs. temperature is plotted in Fig. \ref{fig_EOS}(d). The non-zero peak implies the dual quantum field theory, i.e. QCD,  is not conformal anymore. Similarly, around the phase transition temperature, the trace anomaly jumps from almost zero to a finite value.

\section{Chiral Symmetry Breaking}
In the previous section, we have constructed a hQCD model from a EMS system. A family of analytic black hole solutions was obtained in Eqs. (\ref{phip-A}-\ref{V-A}). We further studied the thermodynamic properties of the background at finite temperatures and chemical potentials, and investigated the phase transition between black holes in the bulk space-time.

In this section, we will study $\chi_{SB}$ in QCD theory that has been studied numerically in several hQCD models \cite{1303.6929,1610.09814,1810.07019}. In this work, we are going to use an analytic method, the matching method, to study $\chi_{SB}$ in our hQCD model.

It is well known that the quark condensation is the order parameter for $\chi_{SB}$. To study the quark condensation in hQCD, we consider a 5-dimensional probe composite scalar field $X$ as the dual order parameter in the bulk space-time with the action in Eq. (\ref{SX}), such bulk field acts as an external source of a boundary operator.

In this work, we only consider the 2-flavor case, so that the composite scalar field is a $2\times2$ matrix and can be brought to the following diagonal form,
\begin{equation}
X=\frac{\chi(z)}{2} \left(
                 \begin{array}{clr}
                 1&0\\
                 0&1
                 \end{array}
                 \right).
\end{equation}
By varying the action in Eq. (\ref{SX}), the equation of motion for the scalar field $\chi$ in the black hole background is derived as
\begin{equation}
\chi''(z)+\frac{p(z)}{z}\chi'(z)-\frac{q(z)}{z^2}\chi(z)=0,\label{EOM}
\end{equation}
where
\begin{eqnarray}
p(z) &=& z \left[ 3A_s'(z)-\phi_s'(z)+\frac{g'(z)}{g(z)}\right]-3, \label{p}\\
q(z) &=&\frac{e^{2A_s(z)} m_{\chi}^2}{g(z)}, \label{q}
\end{eqnarray}
are both regular at the boundary $z=0$.

\subsection{Solving the Scalar Field}
Because the solved black hole background is very complicated, it is hard to solve the the equation of motion exactly. In most of the previous literature, the the equation of motion was solved numerically. In this work, we will use the matching method to obtain an approximated analytic solution. We first solve Eq. (\ref{EOM}) near the boundary at $z=0$ and near the horizon at $z=z_H$ respectively, then connect the two asymptotic solutions by smoothly matching them at an intermediate point $z_\epsilon$,
\begin{align}
\chi_H(z_{\epsilon})=\chi_B(z_{\epsilon}),
~\chi_H'(z_{\epsilon})=\chi_B'(z_{\epsilon}), \label{match condition}
\end{align}
where the sub-index $B/H$ in $\chi$ labels the expansion near the boundary/horizon respectively and $z_\epsilon \in [0,z_H]$. Since the functions $p$ and $q$ in Eqs. (\ref{p}-\ref{q}) are analytic at $z = 0$, the equation of motion Eq. (\ref{EOM}) can be solved by using the Frobenius method near the boundary at $z=0$ with the expansion,
\begin{align}
\chi(z)=\sum\limits_{n=0}^{\infty}C_nz^{n+\lambda},~ \lambda \in \R,
\end{align} 
where the parameter $\lambda$ will be determined from indicial equation in the following.

With the above expansion at $z=0$, the equation of motion becomes
\begin{align}
& \sum_{n=0}^{\infty}C_n(n+\lambda)(n+\lambda-1)z^{n+\lambda-2}+p(z)\sum_{n=0}^{\infty}C_n(n+\lambda)z^{n+\lambda-2}-q(z)\sum_{n=0}^{\infty}C_nz^{n+\lambda-2}=0.
\end{align}
At the leading order, $n=0$, the above equation reduces to
\begin{align}
&C_0[\lambda^{2}-\lambda+p(0)\lambda-q(0)]z^{\lambda-2}=0.
\end{align}
For $C_0\ne 0$ and using  $p(0) = -3$ and $q(0) = m_{\chi}^2$, we obtain the indicial equation,
\begin{align}
&\lambda^{2}-4\lambda- m_{\chi}^2=0,
\end{align}
which has two solutions,
\begin{align}
&\lambda_{1,2}=2\pm\sqrt{4+m_{\chi}^2}.
\end{align}
For a 5-dimensional scalar, we choose $m^2_{\chi}=-3$ to saturate the Breitenlohner-Freedman bound, so that the two solutions reduces to $\lambda_{1}=3$ and $\lambda_{2}=1$. The two roots of the indicial equation are consist with the massive scalar field in the representative one \cite{0501128}.

Because $\lambda_{1}-\lambda_{2} \in {\Z}^{+}$, the two series solutions corresponding to $\lambda_{1}$ and $\lambda_{2}$ are not independent each other. However, to obtain the general solution, we need to take the linear combination of the two independent series solutions. By the Frobenius method, the general solution of Eq. (\ref{EOM}) can be expressed as the linear combination of the two independent series solutions $\chi_{1}$ and $\chi_{2}$ as follows,
\begin{equation}
\chi_B(z)=\alpha\chi_1(z)+\beta\chi_2(z), \label{linear}
\end{equation}
where $\alpha$ and $\beta$ are two arbitrary coefficients, and
\begin{align}
\chi_1(z)&=\sum\limits_{n=0}^{\infty}C^{(1)}_nz^{n+3}, \label{chi1}\\ 
\chi_2(z)&=C\ln z\sum\limits_{n=0}^{\infty}C_n^{(1)}z^{n+3}+\sum\limits_{n=0}^{\infty}C^{(2)}_nz^{n+1}, \label{chi2}
\end{align}
are two independent solutions of the Eq. (\ref{EOM}).

To determine the coefficients $C$, $C_n^{(1)}$ and $C_n^{(2)}$ in the expansions Eq. (\ref{chi1}) and Eq. (\ref{chi2}), we substitute the series solutions $\chi_{1}$ and $\chi_{2}$ into the equation of motion Eq. (\ref{EOM}),
\begin{align}
\sum\limits_{n=0}^{\infty}[(n+3+C)(n+2)+C(n+3)]C_n^{(1)}z^{n+1}+p(z)\sum\limits_{n=0}^{\infty}(n+3+C)C_n^{(1)}z^{n+1}-q(z)\sum\limits_{n=0}^{\infty}C_n^{(1)}z^{n+1}&=0,\\ 
\sum\limits_{n=1}^{\infty}n(n+1)C^{(2)}_nz^{n-1}+p(z)\sum\limits_{n=0}^{\infty}(n+1)C^{(2)}_nz^{n-1}-q(z)\sum\limits_{n=0}^{\infty}C^{(2)}_nz^{n-1}
+C\ln z\sum\limits_{n=0}^{\infty}(n+3)(n+2)C_n^{(1)}z^{n+1}& \notag \\
+Cp(z)\ln z\sum\limits_{n=0}^{\infty}(n+3)C_n^{(1)}z^{n+1}
-Cq(z)\ln z\sum\limits_{n=0}^{\infty}C_n^{(1)}z^{n+1}&=0.
\end{align}
Using the above equations, the coefficients $C$, $C_n^{(1)}$ and $C_n^{(2)}$ can be solved order by order. 
 
Up to the fourth order, we obtain
\begin{align}
C&=-\frac{\left[p''(0)-q''(0)\right]+2\left[2p'(0)-q'(0)\right]\left[p'(0)-q'(0)\right]}{4}C^{(2)}_0, \label{C}\\
C^{(1)}_1&=\frac{q'(0)-3p'(0)}{3}C^{(1)}_0,\\
C^{(2)}_1&=[p'(0)-q'(0)]C^{(2)}_0,\\
 C^{(2)}_{3}&=\left(\frac{q'(0)}{3}-p'(0)\right)C^{(2)}_{2} \notag\\
& -\frac{\left[p''(0)-q''(0)\right]+2\left[2p'(0)-q'(0)\right]\left[p'(0)-q'(0)\right]}{4}\frac{9p'(0)-4q'(0)}{9}C^{(1)}_{0}C^{(2)}_{0}, \notag\\
                &+\frac{1}{6}\left[(q''(0)-2p''(0))(p'(0)-q'(0))+\frac{1}{3}(q'''(0)-p'''(0))\right]C^{(2)}_0.\label{C22}
\end{align}
Since the coefficients $\alpha$ and $\beta$ in the linear combination of $\chi_{1}$ and $\chi_{2}$ in Eq. (\ref{linear}) are arbitrary constants, we can set the parameters  $C^{(1)}_0=C^{(2)}_0=1$ and $C^{(2)}_2=0$ in the above equations without loosing generality. 

Finally, the near boundary solution can be expanded as 
\begin{align}
\chi_B(z)=&\beta z+\beta C^{(2)}_1 z^2+\alpha z^3+\beta C z^3 \ln z+(\alpha C^{(1)}_1+\beta C^{(2)}_3 )z^4+\beta CC^{(1)}_1 z^4 \ln z+\cdots \label{chi0}
\end{align}

On the other side, the scalar field $\chi$ can be expanded near the horizon at $z=z_{H}$ as well. The regular condition of the equation of motion at the horizon, 
\begin{equation}
g'(z_H)\chi'(z_H)-\frac{e^{2A_s(z_H)} m_{\chi}^2}{z_H^2}\chi(z_H)=0, \label{regularcd}
\end{equation}
implies that $\chi(z_H) \neq 0$ to ensure that we will not get a trivial solution, i.e., the expansion should start from the zeroth order,
\begin{equation}
\chi_H(z)=\sum\limits_{n=0}^{\infty}D_n(z-z_{H})^{n}. \label{chiH}
\end{equation}
where the coefficients $D_{n}$ depend on the background in Eqs. (\ref{phip-A}-\ref{V-A}). By plugging the expansion Eq. (\ref{chiH}) into the equation of motion Eq. (\ref{EOM}), we can obtain the coefficients $D_{n}$ order by order. For instance, we list the first few coefficients as follows,
\begin{align}
D_{1}=&\frac{m_{\chi}^2 e^{2A_{s}(z_H)}}{z_{H}^2g'(z_H)}D_{0} = d_1 D_0, \label{D1}\\
D_{2}=& -\frac{D_1}{4}\left[ A'_{s}(z_H)-\phi_s'(z_H)+\frac{g''(z_H)}{g'(z_H)}-\frac{1}{z_H}-d_1 \right]=d_2 D_0, \label{D2}\\
D_{3}=&D_1 \left[ \frac{d_1^2}{36}+\frac{d_1}{6}\left(-\frac{1}{6z_H} -\frac{2g''}{3g'}+\frac{A_s'}{6} +\frac{\phi_s'}{2}\right) -\frac{g'''}{18g'}+\frac{g''}{12g'}\left(-\frac{1}{z_H} +\frac{g''}{g'}+A_s' -\phi_s'\right) \right. \notag\\
&+\frac{1}{z_H}\left( \frac{1}{6z_H}-\frac{7 A_s'}{9}+\frac{2\phi_s'}{9} \right) \left.-\frac{2 A_s''}{9}-\frac{2}{9} A_s' \phi_s'+\frac{7}{18}
   A_s'^2+\frac{\phi_s''}{9}+\frac{1}{18} \phi_s'^2\right]=d_3 D_0, \label{D3}
\end{align}
Since the equation of motion is an ordinary homogeneous linear differential equation, we could scale the overall coefficient $D_0$ to an arbitrary constant. In the following of this paper, we set $D_0=1$.

Next, we will match the two asympototic solutions at boundary and the horizon smoothly to get a completed solution in the domain $[0, z_H]$. The smooth matching conditions Eq. (\ref{match condition}) become
\begin{eqnarray}
    \chi_H  (z_{\epsilon}) &=& \alpha \chi_1 (z_{\epsilon}) +\beta \chi_2 (z_{\epsilon}), \label{match1}\\
    \chi'_H  (z_{\epsilon}) &=& \alpha \chi'_1 (z_{\epsilon}) +\beta \chi'_2 (z_{\epsilon}), \label{match2}
\end{eqnarray}
Using the above matching equations, we can solve the parameter $\alpha$ as
\begin{equation}
\alpha = \frac{\chi_H(z_{\epsilon}) \chi'_2(z_{\epsilon})-\chi'_H(z_{\epsilon}) \chi_2(z_{\epsilon})}{\chi_1(z_{\epsilon}) \chi'_2(z_{\epsilon})-\chi'_1(z_{\epsilon}) \chi_2(z_{\epsilon})}, \label{alpha eq}
\end{equation}
and the dynamical matching point $ z_{\epsilon}$ can be determined from the equation  
\begin{equation}
    \frac{\chi_H(z_{\epsilon}) \chi'_1(z_{\epsilon})-\chi'_H(z_{\epsilon}) \chi_1(z_{\epsilon})}{\chi_2(z_{\epsilon}) \chi'_1(z_{\epsilon})-\chi'_2(z_{\epsilon}) \chi_1(z_{\epsilon})}=\beta, \label{zepsilon}
\end{equation}
where the parameter $\beta$ is a constant associating to the quark mass. 

We expand $\chi_{B}$ to the fourth order and $\chi_{H}$ to the third order in the following explicit form,
\begin{align}
\chi_1(z)&  =z^3+C^{(1)}_{1}z^4,\\
\chi_2(z)&=z+C^{(2)}_{1}z^2+C z^3\ln z+C^{(2)}_{3}z^4+CC^{(1)}_{1}z^4 \ln z,\\
\chi_H(z)&=1+D_1(z-z_{H})+D_2(z-z_{H})^2+D_3(z-z_{H})^3,
\end{align}
where the coefficients $C$'s and $D$'s are calculated in Eqs. (\ref{C}-\ref{C22}) and Eqs. (\ref{D1}-\ref{D3}).

\begin{figure}[t!]
\begin{center}
\includegraphics[
height=2in, width=2.8in]
{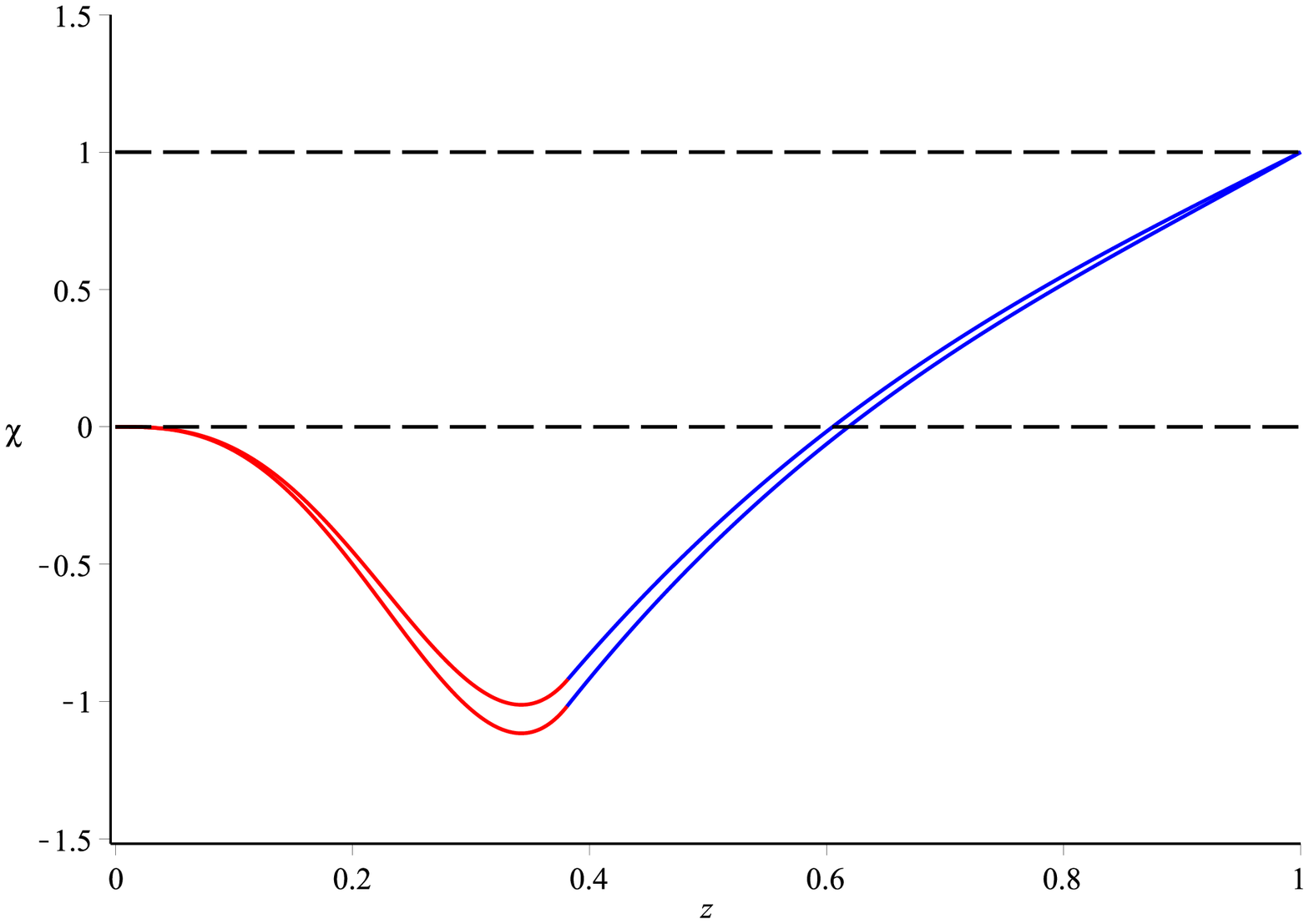}
\includegraphics[
height=2in, width=2.8in]
{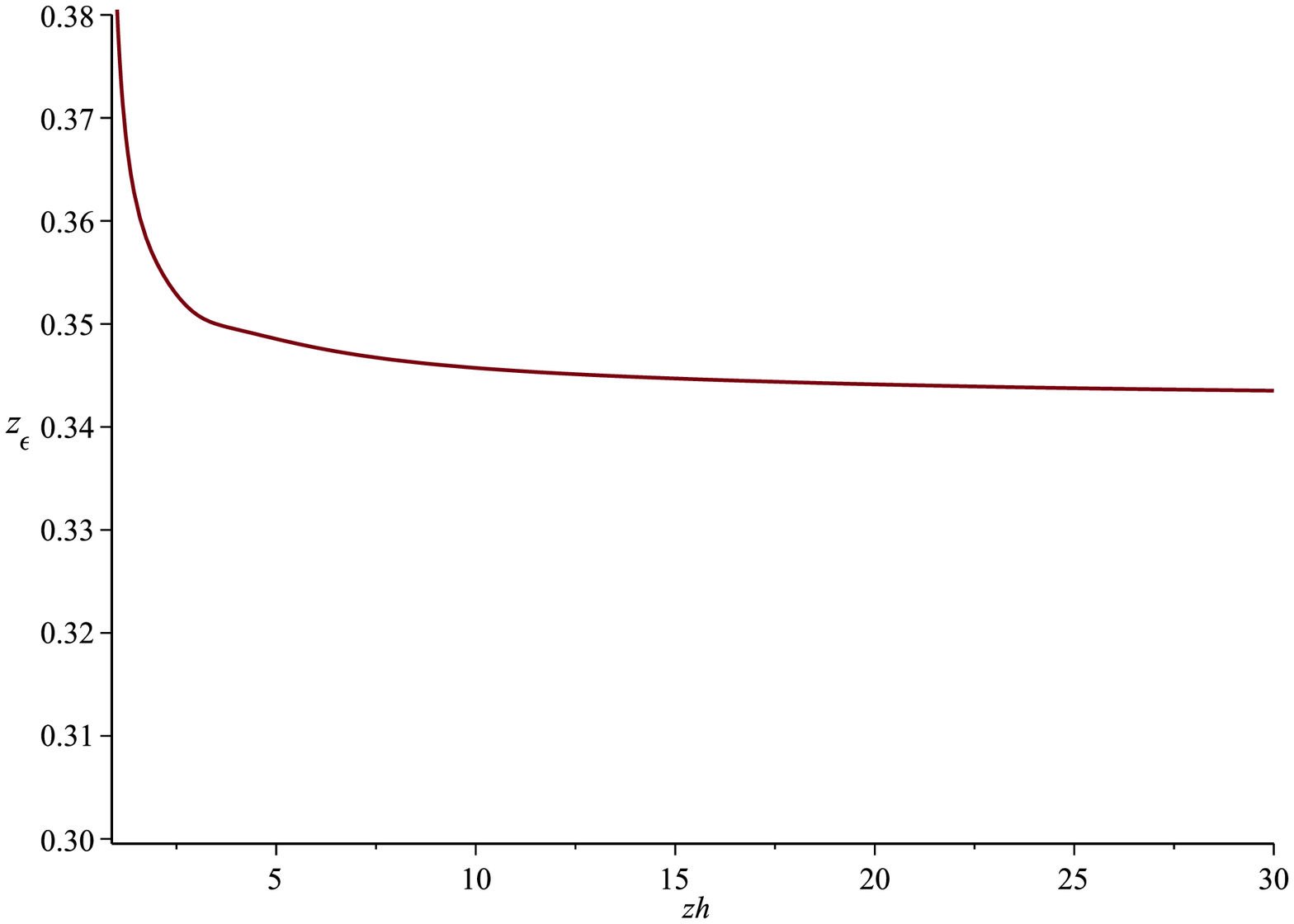}
\vskip -0.05cm \hskip 1 cm (a) \hskip 7 cm (b)
\end{center}
\caption{ (a) The smoothly matching solution for $\mu=0$ (top) and $\mu=1$ (down) at horizon $z_H=1$. The red branches represent asymptotic solutions $\chi_B$ near the boundary $z=0$ and the blue branches are asymptotic solutions $\chi_H$ near the horizon $z=z_{H}$. The two branches are smoothly connected at the matching point $z_{\epsilon}$ which is obtained by solving Eq. (\ref{zepsilon}). (b) The matching points vs. horizons. The matching point is saturated for large $z_H$, i.e. the low temperature. }\label{solution}
\end{figure}

It is worth while to emphasize that the matching point $z_{\epsilon}$ is not a fixed constant but a dynamical variable depending on the temperature, the chemical potential as well as the quark mass through Eq. (\ref{zepsilon}). (Without loss of generality, the $z_{\epsilon}$ is linear with the black hole horizon because of the dimensional analysis, so we anchor the $z_{\epsilon}=\epsilon z_H$.) As an example, the smoothly matched solution for $z_H=1$ and $\mu=0, 1$ are plotted in Fig. \ref{solution}(a). The red part is the near boundary solution with boundary condition $\chi(0)=0$, and the blue part is the near horizon solution with the boundary condition $\chi(z_H)=1$. The two asymptotic solutions are smoothly connected at the matching point $z_{\epsilon}\sim 0.381$. For different chemical potentials, the matching point is almost saturated in a constant as showed in Fig. \ref{solution}(a).

In Fig. \ref{solution}(b), we plot the matching points for different horizons, or different temperatures. The matching point varies slowly for large $z_H$, and approaches to a saturated value for large $z_H$, i.e. low temperature, as expected since quarks are in a steady condensation state. For small $z_H$ , i.e, high temperature, the matching point increases.

Once we fixed the matching point $z_{\epsilon}$, the parameter $\alpha$ can be obtained from Eq. (\ref{alpha eq}).
\subsection{Quark Condensation}
We have obtained an approximate analytic solution of the probe scalar field $\chi$ by using the matching method in the last section. By holographic correspondence, the 5-dimensional massive scalar field $\chi$ in an asymptotic $AdS_5$ space-time can be expanded near the boundary as,
\begin{equation}
\chi_B\left( z \right) =\alpha\chi_1(z)+\beta\chi_2(z)=\beta z+...+\alpha z^3+\cdots,
\end{equation}
where the linear term is the leading power of $\chi_2$ that dominates near the UV boundary, and the cubic term is the leading power of $\chi_1$. Notice that, depending on the details of the bulk space-time, there might be other terms in the expansion between these two powers that are not the leading power in either $\chi_1$ and $\chi_2$.

In the system we are considering, the holographic dictionary claims that the linear term in the near boundary expansion corresponds to the source of the dual operator and the cubic term represents the response which is triggered by the source term. In the hQCD treatment, quark mass plays the role of the source term, and the quark condensation is the effective response,
\begin{equation}
\chi_B\left( z \right) = m_{q}\zeta z + ... + \frac{\Sigma}{\zeta}z^3+ \cdots, \label{boundary chi}
\end{equation}
where $\zeta =N_c/2\pi N_f^{1/2}$ is a constant\footnote{We take $N_c=N_f=3$ in this work.}, $m_q \simeq 3MeV$ is the current quark mass, and $\Sigma=\langle \bar{\psi}\psi \rangle $ represents the quark condensate. 

Comparing Eq. (\ref{chi0}) and Eq. (\ref{boundary chi}), we have
\begin{equation}
\Sigma=\alpha\zeta, ~m_{q}=\frac{\beta}{\zeta}. \label{ab}
\end{equation}
Since we have obtained the parameter $\alpha$ in Eq. (\ref{alpha eq}), the quark condensate $\Sigma$ can be calculated once the quark mass is given.
\begin{figure}[t!]
\begin{center}
\includegraphics[
height=3in, width=4.5in]
{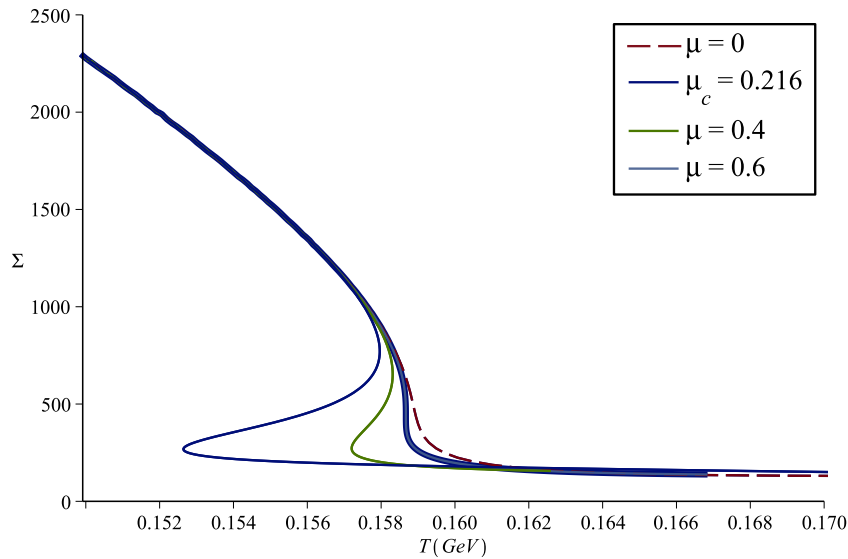}
\end{center}
\caption{ Quark condensation vs. temperature at different chemical potentials. At low temperature, the non-zero condensation implies the chiral symmetry breaking. While at high temperature, the condensation approaches to almost zero that implies the chiral symmetry restore. For $\mu<\mu_c$, the quark condensation is a monotonous function of temperature, which becomes multi-valued for $\mu>\mu_c$. } \label{fig_condensation}
\end{figure}

The quark condensation vs. temperature for different chemical potentials is plotted in Fig. \ref{fig_condensation}. At high temperature, the quark condensation approaches to a small value indicating that the QCD is in the (almost) chiral symmetry phase. Notice that, since we consider the finite quark mass, we do not expect the exact chiral symmetry, so that the quark condensation at high temperature is small but not exact zero. While at low temperature, the quark condensation becomes non-zero implying the chiral symmetry is breaking.

\begin{figure}[t!]
\begin{center}
\hskip -1cm
\includegraphics[
height=1.8in, width=2.8in]
{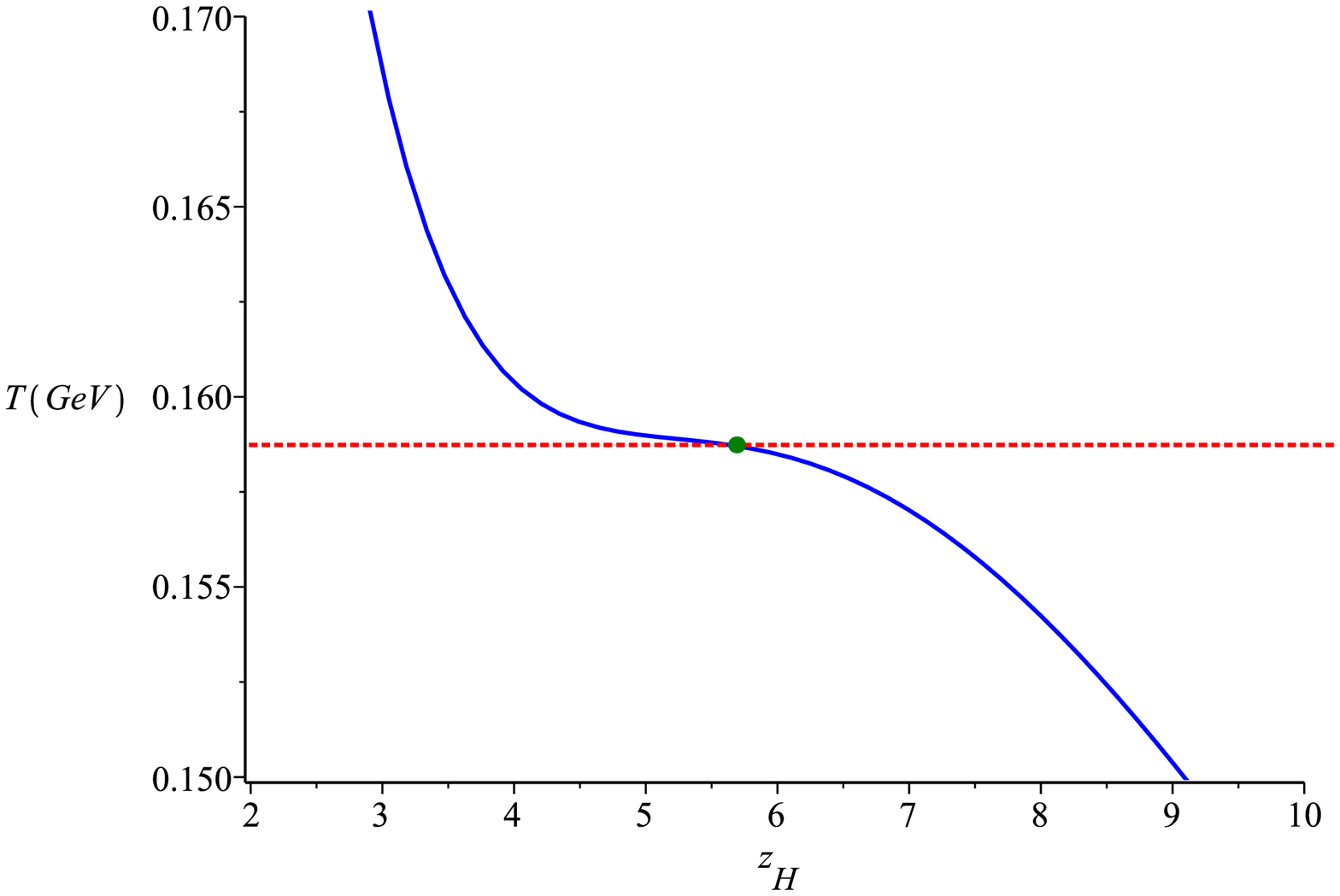}
\includegraphics[
height=1.8in, width=2.8in]
{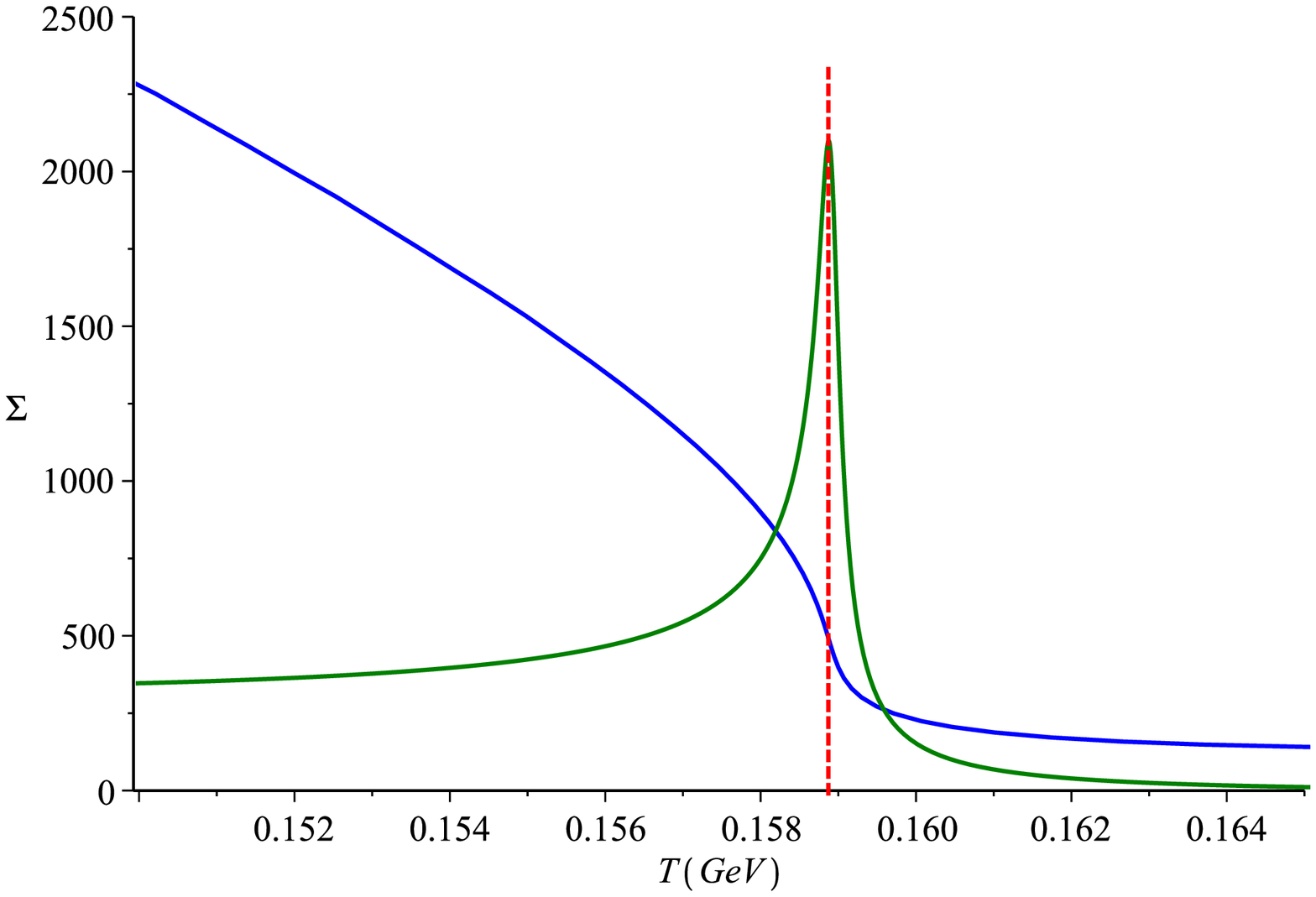}
\vskip -0.05cm \hskip 1 cm (a) \hskip 7 cm (b) \\
 \hskip -1 cm
\includegraphics[
height=1.8in, width=2.8in]
{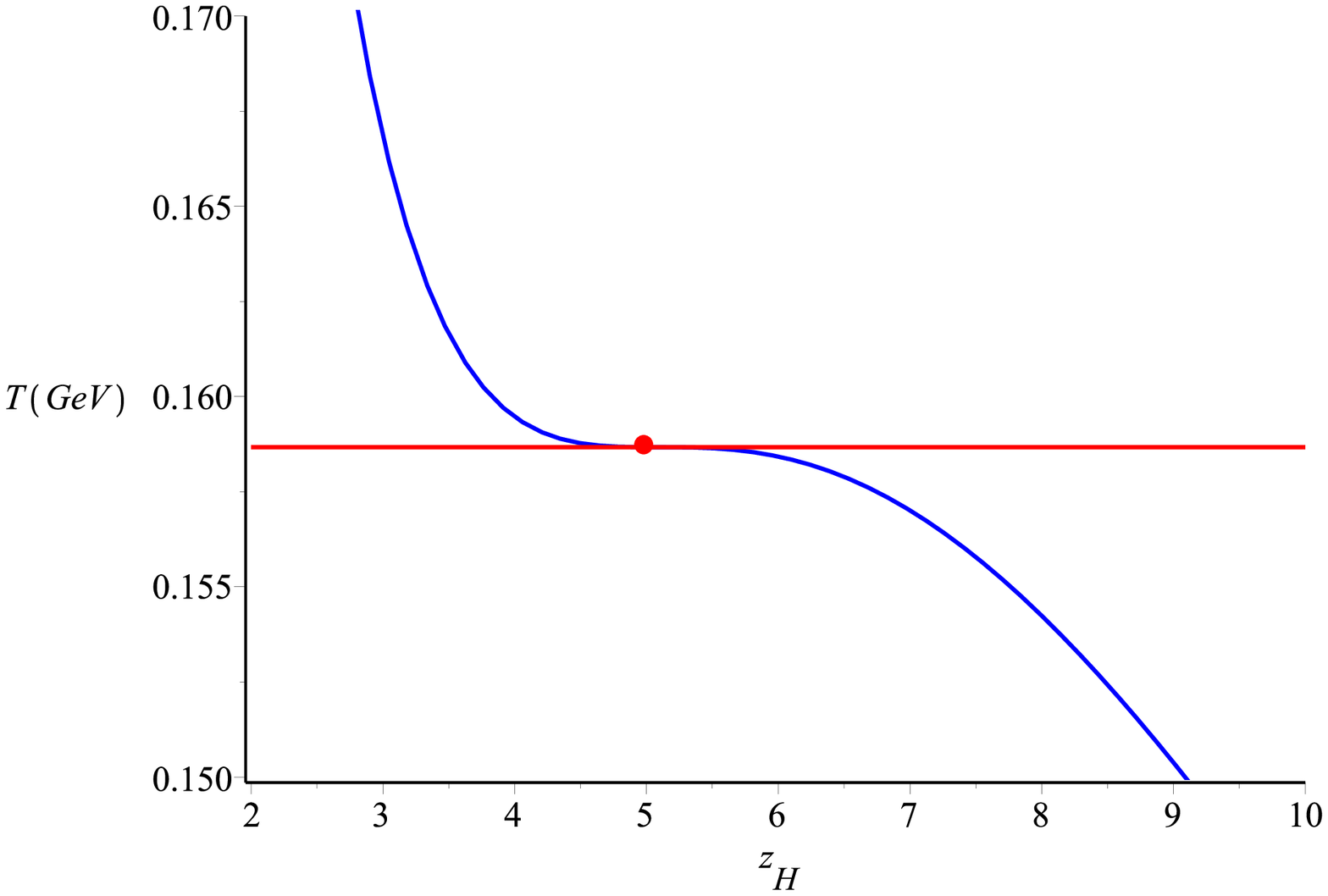}
\includegraphics[
height=1.8in, width=2.8in]
{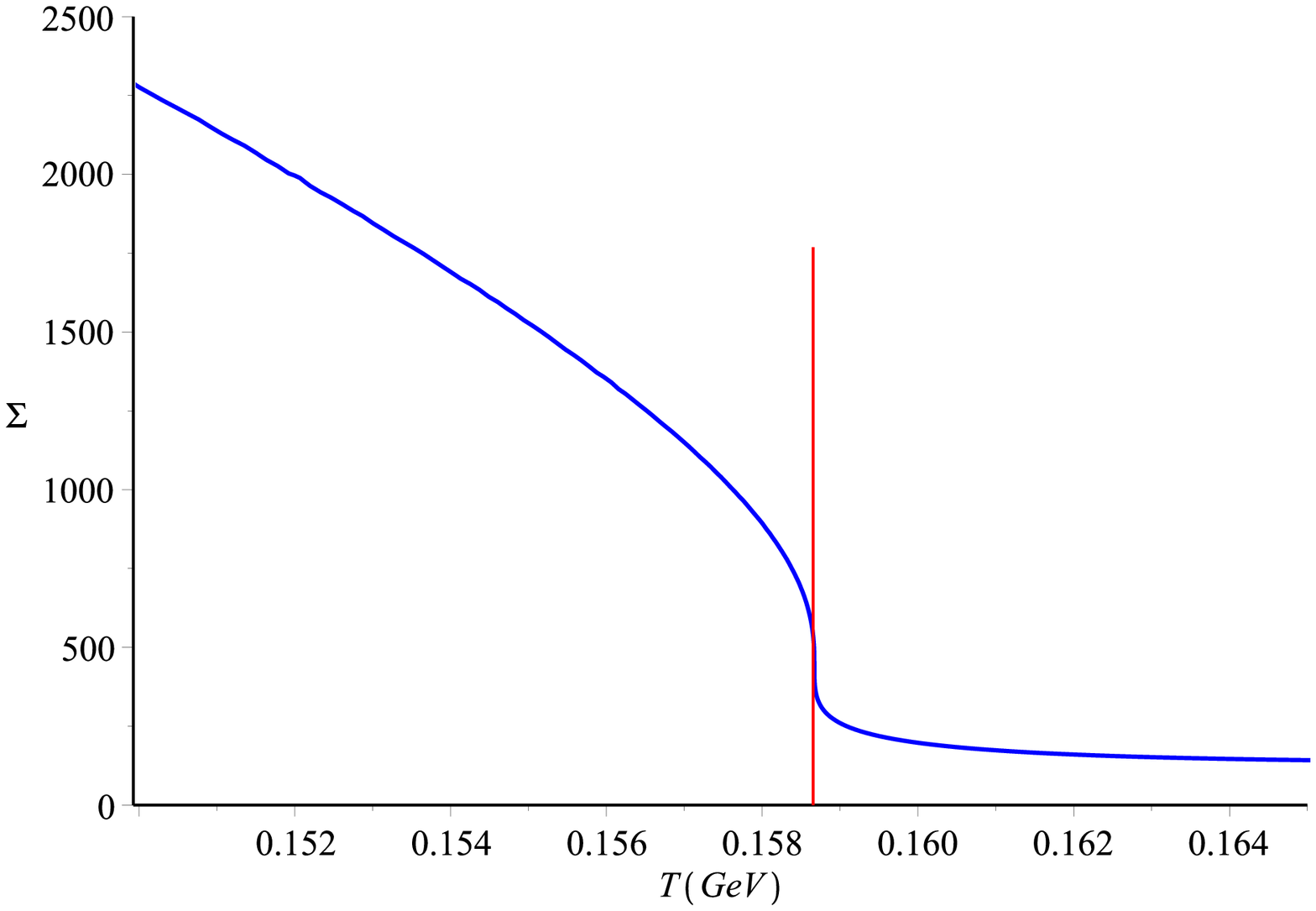}
\vskip -0.05cm \hskip 1 cm (c) \hskip 7 cm (d) \\
 \hskip -1 cm
\includegraphics[
height=1.8in, width=2.8in]
{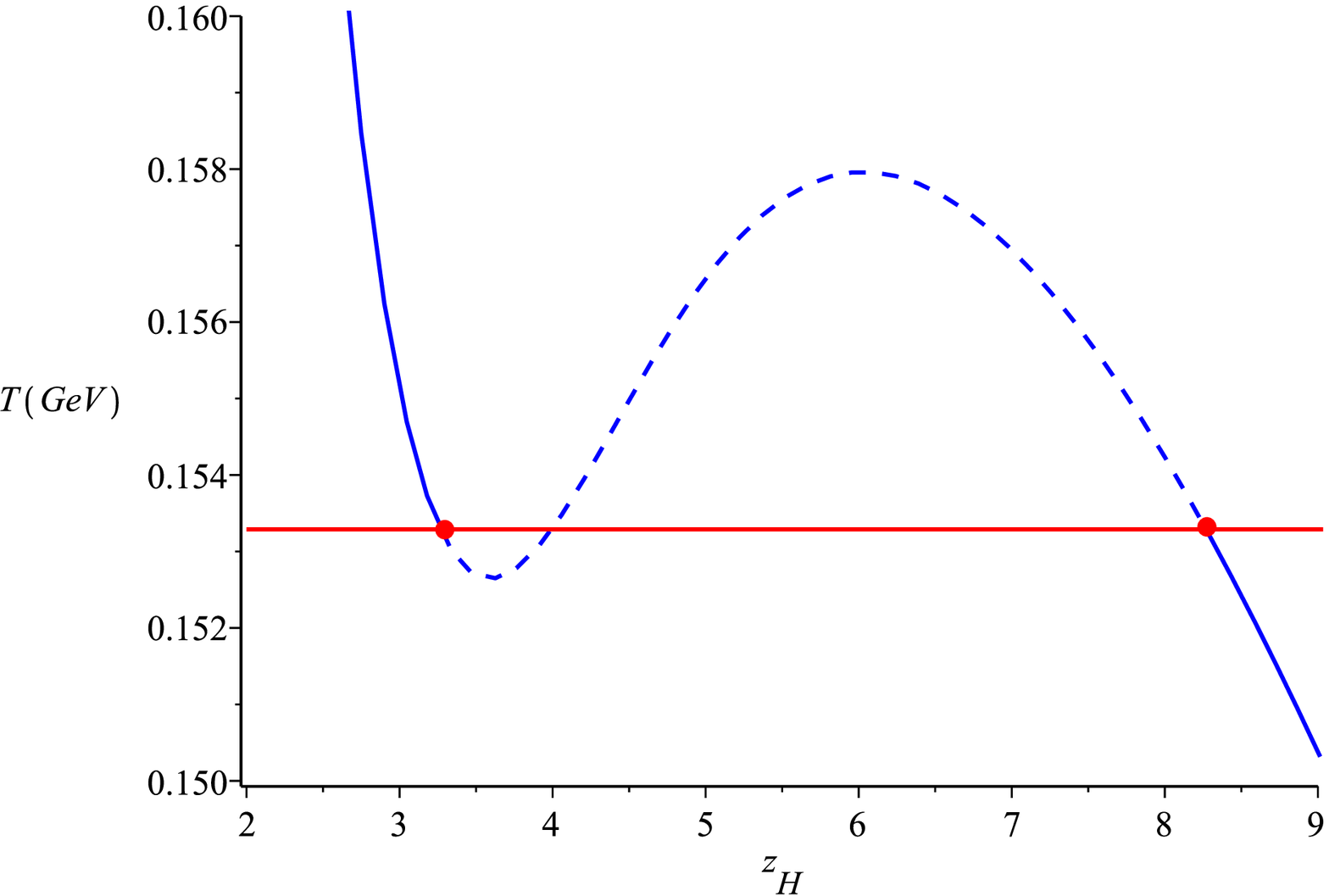}
\includegraphics[
height=1.8in, width=2.8in]
{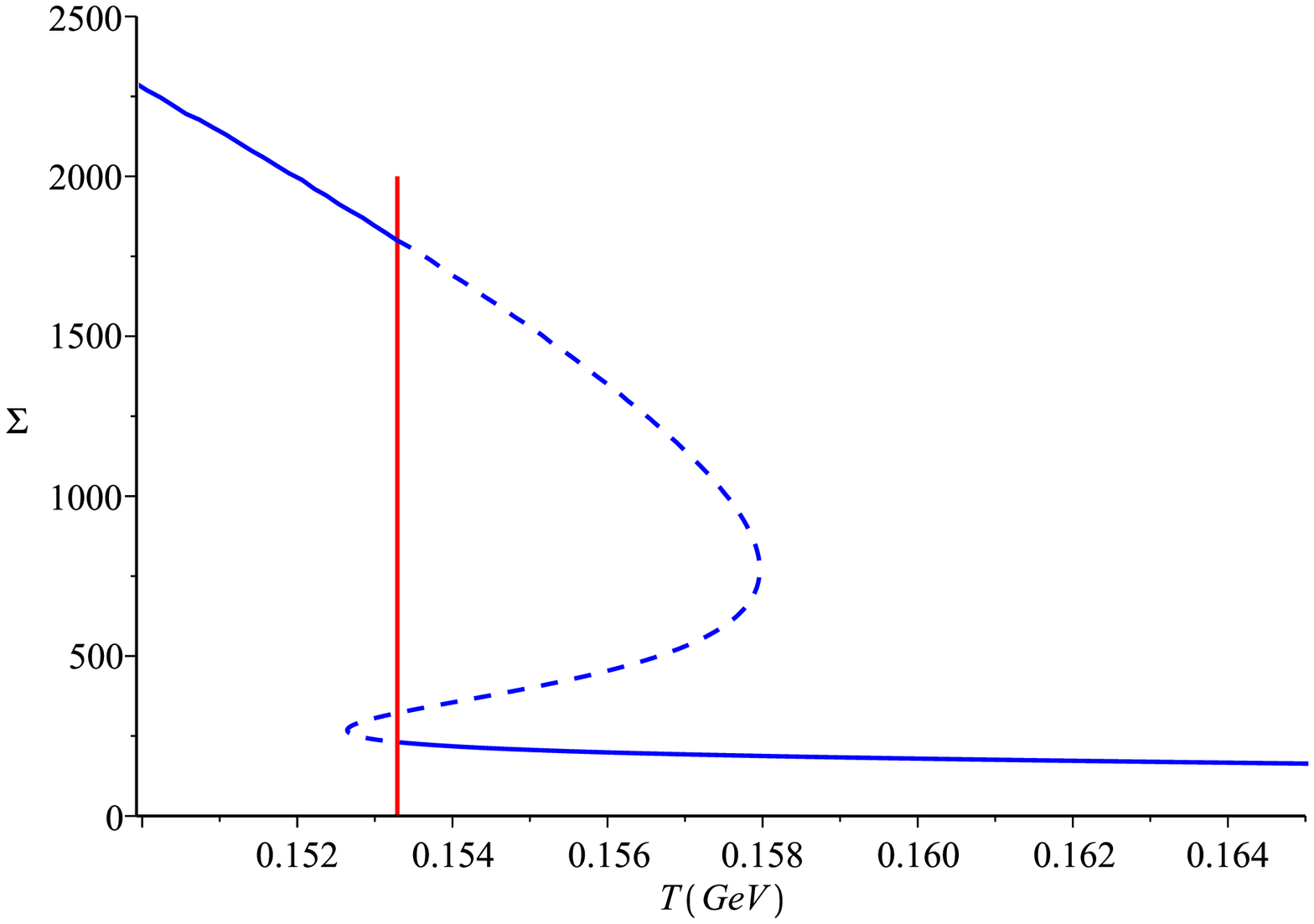}
\vskip -0.05cm \hskip 1 cm (e) \hskip 7 cm (f) 
\end{center}
\caption{ The black hole temperature vs. horizon and the quark condensation vs. temperature for chemical potential at $\mu=0$ (a,b), $\mu=\mu_c$ (c,d) and $\mu=0.6$ (e,f). At $\mu=0$, both black hole temperature and quark condensation are monotonous. The green line represents the derivative of the quark condensation with the transition temperature labeled by the red dashed line at its peak. At $\mu=\mu_c$, there is a saddle point at the critical temperature which causes the derivative of condensation divergent. For $\mu>\mu_c$, the multi-valued behavior implies a black hole phase transition between two horizons labelled by the red dots in (e). The dashed blue curves indicate the unstable region which is bypassed by the black hole phase transition. The bypass mechanism forces the quark condensation phase transition takes place at the same temperature of black hole phase transition in (f).} \label{SigmaT_trinity}
\end{figure}

For small chemical potential, the quark condensation is monotonic decreasing with temperature growing. The critical temperature of quark condensation can be determined by the temperature with the maximum changing rate as showed in Fig. \ref{SigmaT_trinity}(b).
On the other hand, for large enough chemical potential, the quark condensation becomes multi-valued with a local maximum and a local minimum temperatures. The multi-valued behavior implies that a first order phase transition would happen between the two local extreme temperatures. In previous section, we have studied the black hole phase transition in the bulk space-time. At the transition temperature, the horizon jumps between a small black hole and a large one, which affects $\chi_{SB}$ seriously. To obtain the phase diagram of $\chi_{SB}$, we need to combine the quark condensation with the black hole phase transition in the bulk space-time together. We will explain this effect by examining the phase transitions at some typical chemical potentials, $\mu=0$, $\mu=\mu_c, $ and $\mu>\mu_c$, as plotted in Fig. \ref{SigmaT_trinity}.

At zero chemical potential, $\mu=0$, Fig. \ref{SigmaT_trinity}(a) shows the black hole temperature vs. horizon. As we discussed in previous section, the temperature in this case is monotonic indicating the background transition is a crossover. The quark condensation (blue curve) at zero chemical potential is plotted in Fig. \ref{SigmaT_trinity} (b), that is also monotonic. We determine the crossover temperature by looking for the temperature with its maximum changing rate. The green curve in Fig.\ref{SigmaT_trinity} (b) represents the derivative of the quark condensation. We find that the crossover temperature is around $T\sim 0.159 ~GeV$. The red dashed line in Fig. \ref{SigmaT_trinity}(a, b) represents the crossover transition temperature, and the green dot labels the black hole horizon at the transition temperature.

At the critical chemical potential $\mu_c \sim ~0.216 ~GeV$, the derivative of the quark condensation diverges at the transition temperature that indicates a second order phase transition. The red lines in Fig. \ref{SigmaT_trinity}(c, d) represents the critical phase transition temperature $T_c\sim 0.159~ GeV$, which is consistent with the lattice simulation result \cite{1701.04325}.

Finally, when the chemical potential is large enough, i.e. $\mu>\mu_c$. Both the temperature and the quark condensation become multi-valued. We plot temperature vs. horizon at $\mu=0.6 ~GeV$ in Fig. \ref{SigmaT_trinity}(e). The horizontal red line represents the phase transition temperature between a small black hole and a large one labelled by two red dots. The dashed segment between the two red dots indicates the thermodynamic unstable region where the real physical state can not occur. The quark condensation vs. temperature is plotted in Fig. \ref{SigmaT_trinity}(f). When the black hole phase transition takes place, a black hole suddenly jumps between two horizons, which causes the quark condensate drops to almost zero at the transition temperature. When this happens, the dashed segment of the quark condensation curve is bypassed. This "bypass" mechanism force the quark condensation to take place at the same temperature of the black hole phase transition no matter what temperature the $\chi_{SB}$ is supposed to take place.

\begin{figure}[t!]
\begin{center}
\includegraphics[
height=3in, width=4.5in]
{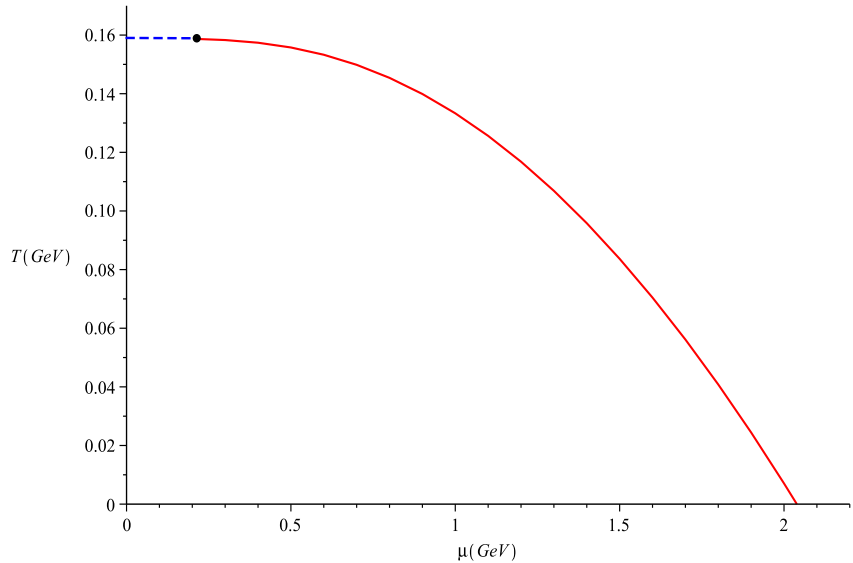}
\end{center}
\caption{ The phase diagram of $\chi_{SB}$, in which the second order phase transition point is identified as the CEP of the phase transition. } \label{Tmu}
\end{figure}

By combining the quark condensation and black hole phase transition together, we finally determined the phase diagram of $\chi_{SB}$, as plotted in Fig. \ref{Tmu}. The second order phase transition point is identified as the critical end point (CEP), whose value is consistent with the recent result by lattice QCD in \cite{1701.04325}. For small chemical potential $\mu<\mu_c$, the transition is crossover instead of phase transition, because the order parameters change smoothly in this region. While for large chemical potential $\mu>\mu_c$, the bypass mechanism forces the quark condensation phase transition to take place at the same transition temperature as the black hole phase transition. 

\section{Conclusion}
In this paper, we studied a holographic QCD model to understand the mechanism of the chiral phase transition by using the Einstein-Maxwell-Dilaton system with potential reconstruction approach. A family of analytic black hole solutions is obtained. To include meson fields in QCD, we added a probe gauge field on the 5-dimensional backgrounds and studied the linear Regge spectrum of mesons.

In the bulk space-time, we found a black hole phase transition between black holes with different sizes for large enough chemical potential by calculating the free energy. To further understand the phase structure of the thermal background we calculated the equations of states, including the speed of sound, the specific heat and the trace anomaly. For large chemical potential $\mu>\mu_c$, the black hole phase transition is first order. At the critical end point $(\mu_c,T_c) \simeq (0.216,0.159)$ $GeV$, the phase transition becomes second order. For small chemical potential $\mu<\mu_c$, the phase transition reduces a crossover as expected from Lattice QCD simulations.

The chiral symmetry breaking was studied by considering a probing composite scalar field which plays the role of the composite operator in the dual QCD theory. The massive scalar field implies that the equation of motion can be solved by two independent series solutions. The leading order of the source solution is linear in $z$, and the leading order of the response solution is cubic in $z$. By holographic correspondence, the coefficient of the leading term of the response solution is identified to the quark condensation which is triggered by the coefficient of the leading term of the source solution, the quark mass. 

To obtain an analytic solution for the scalar field, we solved the equation of motion near the boundary and the horizon respectively, and smoothly matched the two branches at a matching point. We found that the matching point is a dynamical quantity which depends on both the temperature and the chemical potential. Our result shows that the quark condensation approaches to zero in the high temperature limit, where the chiral symmetry is stored. However, in the low temperature region the quark condensation is non-zero, implying that the chiral symmetry is broken. For small chemical potential $\mu<\mu_c$, the quark condensation is a monotonous function of the temperature, we identifies the peak of its changing rate as the transition temperature of the crossover. For large chemical potential $\mu>\mu_c$, the multi-valued behavior of the quark condensation implies that there is a first order phase transition. 

To obtain the correct phase diagram of the chiral symmetry breaking, we need to consider the effect of the black hole phase transition in the bulk space-time on the quark condensation from the probe scalar field. When the black hole phase transition takes place, a black hole jumps between two horizons and bypasses the medium unstable region. This bypass mechanism forces the quark condensation phase transition to take place at the same transition temperature as the black hole phase transition. 

As we mentioned in the introduction, the mass dependent of the phase transition behavior in QCD is a very important and unsolved problem. In the current work, we fitted the Regge trajectory from the masses of $\rho$ mesons. However, the quark condensation is a well-defined order parameter only in the chiral limit, $m_q \to 0$. In addition, the chiral susceptibility is defined as $\frac{\partial \Sigma}{\partial m_q}$, and lattice QCD simulations showed that the quark mass does affect the phase structure. To fully understand the chiral symmetry breaking in QCD, it is necessary to impose the variable quark mass. Thus we need to generalize the parameters $a$ and $b$ in the warped factor Eq. (\ref{ansatz_A}) to be functions of $m_{q}$. The values of $a$ and $b$ determine the transition temperature at $\mu=0$ and the locations of the critical point in a complex way. We have realized that the transition would become first order in the chiral limit if the parameter $a$ decreases with $m_{q}$. However, the concrete form of the function $a\left( m_{q} \right)$ is not completely determined yet. We will leave this issue in the future.

\section*{Acknowledgements}
We would like to thank Chiang-Mei Chen, Song He, Danning Li, Xuanting Ji, Wen-Yu Wen, Xin-Meng Wu, Zhongshan Xu for useful discussions. This work of Y.Y is supported by the Ministry of Science and Technology (MOST 106-2112-M-009 -005 -MY3) and National Center for Theoretical Science, Taiwan. The work of PHY was supported by the University of Chinese Academy of Sciences.

\end{document}